\documentclass{math_crypt_v06_modified}

\title{Minimal weight expansions in Pisot bases}

\authorone{Christiane Frougny}
\addressone{LIAFA, CNRS UMR 7089, and Universit\'{e} Paris 8}
\countryone{France}
\emailone{christiane.frougny@liafa.jussieu.fr}

\authortwo{Wolfgang Steiner}
\addresstwo{LIAFA, CNRS UMR 7089, Universit\'e Paris Diderot -- Paris 7,
Case 7014, 75205 Paris Cedex 13}
\countrytwo{France}
\emailtwo{steiner@liafa.jussieu.fr}

\researchsupported{This research was supported by the French Agence Nationale de la Recherche, grant ANR-JCJC06-134288 ``DyCoNum''.} 

\firstpage{1}                                %
\doi{}                                       %
\communicated{}                              %
\received{}                                  %
\revised{}                                   %

\abstract{For applications to cryptography, it is important to represent numbers with a small number of non-zero digits (Hamming weight) or with small absolute sum of digits. 
The problem of finding representations with minimal weight has been solved for integer bases, e.g. by the non-adjacent form in base~2.
In this paper, we consider numeration systems with respect to real bases $\beta$ which are Pisot numbers and prove that the expansions with minimal absolute sum of digits are recognizable by finite automata.
When $\beta$ is the Golden Ratio, the Tribonacci number or the smallest Pisot number, we determine expansions with minimal number of digits $\pm1$ and give  explicitely the finite automata recognizing all these expansions.
The average weight is lower than for the non-adjacent form. }

                                        

\def\decdot{\raisebox{0.1ex}{\textbf{.}}}

\begin{document}

\section{Introduction}
Let $A$ be a set of (integer) digits and $x=x_1x_2\cdots x_n$ be a word with letters $x_j$ in $A$.  
The \emph{weight} of $x$ is the \emph{absolute sum of digits} 
$\|x\|= \sum_{j=1}^n |x_j|$.
The \emph{Hamming weight} of $x$ is the number of non-zero digits in $x$.
Of course, when $A\subseteq \{-1,0,1\}$, the two definitions coincide.

Expansions of minimal weight in integer bases $\beta$ have been studied extensively. 
When $\beta=2$, it is known since Booth~\cite{Booth51} and Reitwiesner~\cite{Reitwiesner60} how to obtain such an expansion with the digit set $\{-1,0,1\}$.
The well-known non-adjacent form (NAF) is a particular expansion of minimal weight with the property that the non-zero digits are isolated.
It has many applications to cryptography, see in particular~\cite{MorainOlivos90,JoyeTymen01,MuirStinson06}.
Other expansions of minimal weight in integer base are studied in~\cite{GrabnerHeuberger06,HeubergerProdinger06}.
Ergodic properties of signed binary expansions are established in~\cite{DKL06}.

Non-standard number systems --- where the base is not an integer --- have been studied from various points of view.
Expansions in a real non-integral base $\beta >1$ have been introduced by R\'enyi~\cite{Renyi57} and studied initially by Parry~\cite{Parry60}. 
Number theoretic transforms where numbers are represented in base the Golden Ratio have been introduced in~\cite{DCD95} for application to signal processing and fast convolution.
Fibonacci representations have been used in~\cite{Meloni07} to design exponentiation algorithms based on addition chains.
Recently, the investigation of minimal weight expansions has been extended to the Fibonacci numeration system by Heuberger~\cite{Heuberger04}, who gave an equivalent to the NAF.  
Solinas~\cite{Solinas00} has shown how to represent a scalar in a complex base $\tau$ related to Koblitz curves, and has given a $\tau$-NAF form, and the Hamming weight of these representations has been studied in~\cite{EbeidHasan07}.

\smallskip
In this paper, we study expansions in a real base $\beta>1$ which is not an 
integer. 
Any number $z$ in the interval $[0,1)$ has a so-called 
\emph{greedy $\beta$-expansion} given by the $\beta$-transformation 
$\tau_\beta$, which relies on a greedy algorithm: 
let $\tau_\beta(z)=\beta z-\lfloor\beta z\rfloor$ and define, for $j \ge 1$, 
$x_j=\lfloor \beta \tau_\beta^{j-1}(z) \rfloor$.
Then $z=\sum_{j=1}^\infty x_j\beta^{-j}$, where the $x_j$'s are integer digits 
in the alphabet $\{0,1,\ldots,\lfloor\beta\rfloor\}$.
We write $z=\decdot x_1x_2 \cdots$.
If there exists a $n$ such that $x_j=0$ for all $j>n$, the expansion is said to 
be \emph{finite} and we write $z=\decdot x_1x_2\cdots x_n$.
By shifting, any non-negative real number has a greedy $\beta$-expansion:
If $z\in[0,\beta^k)$, $k\ge0$, and $\beta^{-k}z=\decdot x_1x_2\cdots$,
then $z=x_1\cdots x_k\decdot x_{k+1}x_{k+2}\cdots$.

We consider the sequences of digits $x_1x_2\cdots$ as words.
Since we want to minimize the weight, we are only interested in finite words 
$x=x_1x_2\cdots x_n$, but we allow a priori arbitrary digits $x_j$ in 
$\mathbb Z$.
The corresponding set of numbers $z=\decdot x_1x_2\cdots x_n$ is therefore 
$\mathbb Z[\beta^{-1}]$. 
Note that we do not require that the greedy $\beta$-expansion of every 
$z\in\mathbb Z[\beta^{-1}]\cap[0,1)$ is finite, although this property (F) 
holds for the three numbers $\beta$ studied in Sections~\ref{GR} to~\ref{sp}, 
see~\cite{FrougnySolomyak92,Akiyama00}.

The set of finite words with letters in an alphabet $A$ is denoted by $A^*$, as
usual.
We define a relation on words $x=x_1x_2\cdots x_n\in\mathbb Z^*$,
$y=y_1y_2\cdots y_m\in\mathbb Z^*$ by 
$$
x\sim_\beta y\quad\mbox{ if and only if }\quad
\decdot x_1x_2\cdots x_n=\beta^k\times\decdot y_1y_2\cdots y_m
\mbox{ for some }k\in\mathbb Z.
$$
A word $x\in\mathbb Z^*$ is said to be \emph{$\beta$-heavy} if there exists 
$y\in\mathbb Z^*$ such that $x\sim_\beta y$ and $\|y\|<\|x\|$.
We say that $y$ is \emph{$\beta$-lighter} than $x$.
This means that an appropriate shift of $y$ provides a $\beta$-expansion of the 
number $\decdot x_1x_2\cdots x_n$ with smaller absolute sum of digits than 
$\|x\|$.
If $x$ is not $\beta$-heavy, then we call $x$ a 
\emph{$\beta$-expansion of minimal weight}.
It is easy to see that every word containing a $\beta$-heavy factor is 
$\beta$-heavy.
Therefore we can restrict our attention to \emph{strictly $\beta$-heavy} words
$x=x_1\cdots x_n\in\mathbb Z^*$, which means that $x$ is $\beta$-heavy, and 
$x_1\cdots x_{n-1}$ and $x_2\cdots x_n$ are not $\beta$-heavy.

\smallskip
In the following, we consider real bases $\beta$ satisfying the condition
$$
(\mathrm D_B):\ \mbox{there exists a word }b\in\{1-B,\ldots,B-1\}^* \\ 
\mbox{ such that } B\sim_\beta b\mbox{ and }\|b\|\le B
$$
for some positive integer $B$.
Corollary~\ref{cARS} and Remark~\ref{rB} show that every class of words (with respect to~$\sim_\beta$) contains a $\beta$-expansion of minimal weight with digits in $\{1-B,\ldots,B-1\}$ if and only if $\beta$ satisfies ($\mathrm D_B$).

If $\beta$ is a Pisot number, i.e., an algebraic integer greater than~1 such that all the other roots of its minimal polynomial are in modulus less than one, then it satisfies ($\mathrm D_B$) for some $B>0$ by Proposition~\ref{existsB}.
The contrary is not true: 
There exist algebraic integers $\beta>1$ satsfying ($\mathrm D_B$) which are not Pisot, e.g. the positive solution of $\beta^4=2\beta+1$ is not a Pisot number but satisfies ($\mathrm D_2$) since $2=1000\decdot(-1)$.
The following example provides a large class of numbers $\beta$ satisfying ($\mathrm D_B$).

\begin{example}\label{ex1}
If $1=\decdot t_1t_2\cdots t_d(t_{d+1})^\omega$ with integers
$t_1\ge t_2\ge\cdots\ge t_d>t_{d+1}\ge0$, then $\beta$ satisfies ($\mathrm D_B$) with $B=t_1+1=\lfloor\beta\rfloor+1$, since 
$$
\beta^{d+1}-t_1\beta^d-\cdots-t_d\beta-t_{d+1}=\frac{t_{d+1}}{\beta-1}=
\beta^d-t_1\beta^{d-1}-\cdots-t_d
$$ 
and thus 
$$
\beta^{d+1}-(1+t_1)\beta^d+(t_1-t_2)\beta^{d-1}+\cdots+(t_{d-1}-t_d)\beta+
(t_d-t_{d+1})=0.
$$
\end{example}

Recall that the set of greedy $\beta$-expansions is recognizable by a finite 
automaton when $\beta$ is a Pisot number~\cite{Bertrand77}.
In this work, we prove that the set of all $\beta$-expansions of minimal weight is recognized by a finite automaton when $\beta$ is a Pisot number.

We then consider particular Pisot numbers satisfying ($\mathrm D_2$) which have been extensively studied from various points of view. 
When $\beta$ is the Golden Ratio, we construct a finite transducer which gives,
for a strictly $\beta$-heavy word as input, a $\beta$-lighter word as output.
Similarly to the Non-Adjacent Form in base 2, we define a particular unique
expansion of minimal weight avoiding a certain given set of factors.
We show that there is a finite transducer which converts all words 
of minimal weight
into these expansions avoiding these factors. 
{}From these transducers, we derive the minimal automaton recognizing the set
of $\beta$-expansions of minimal weight in $\{-1,0,1\}^*$.
We give a branching transformation which provides all $\beta$-expansions of 
minimal weight in $\{-1,0,1\}^*$ of a given $z\in\mathbb Z[\beta^{-1}]$.
Similar results are obtained for the representation of integers in the 
Fibonacci numeration system.
The average weight of expansions of the numbers $-M,\ldots,M$ is 
$\frac15\log_\beta M$, which means that typically only every fifth digit is 
non-zero.
Note that the corresponding value for $2$-expansions of minimal weight is 
$\frac13\log_2 M$, see~\cite{ArnoWheeler93,Bosma07}, and that
$\frac15\log_\beta M\approx0.288\log_2 M$.

We obtain similar results for the case where $\beta$ is the so-called 
\emph{Tribonacci number}, which satisfies $\beta^3=\beta^2+\beta+1$ 
($\beta\approx1.839$), and the corresponding representations for integers.
In this case, the average weight is $\frac{\beta^3}{\beta^5+1}\log_\beta M
\approx 0.282\log_\beta M\approx 0.321\log_2 M$.

Finally we consider the smallest Pisot number, $\beta^3=\beta+1$
($\beta\approx1.325$), which provides representations of integers with even
lower weight than the Fibonacci numeration system:
$\frac1{7+2\beta^2}\log_\beta M\approx 0.095\log_\beta M\approx 0.234\log_2 M$.

Since the proof techniques for the Tribonacci number and the smallest Pisot number are quite similar to the Golden Ratio case (but more complicated), some parts of the proofs are not contained in the final version of this paper.
The interested reader can find them in~\cite{FrougnySteiner08}.

\section{Preliminaries}

A finite sequence of elements of a set $A$ is called a \emph{word}, and 
the set of words on $A$ is the free monoid $A^*$.
The set $A$ is called \emph{alphabet}.
The set of infinite sequences or infinite words on $A$ is denoted by 
$A^\mathbb{N}$. 
Let $v$ be a word of $A^*$, denote by $v^n$ the concatenation of $v$ to itself
$n$ times, and by $v^\omega$  the infinite concatenation $vvv\cdots$.

A finite word $v$ is a \emph{factor} of a (finite or infinite)
word $x$ if there exists $u$ and $w$ such that $x=uvw$.
When $u$ is the empty word, $v$ is a \emph{prefix} of $x$.
The prefix $v$ is \emph{strict} if $v \neq x$.
When $w$ is empty, $v$ is said to be a \emph{suffix} of $x$.

We recall some definitions on automata, see \cite{Eilenberg74} and \cite{Sakarovitch03} for instance.
An \emph{automaton over $A$}, $\mathcal A=(Q,A,E,I,T)$, is a directed graph 
labelled by elements of $A$. 
The set of vertices, traditionally called \emph{states}, is denoted by $Q$, 
$I \subset Q$ is the set of \emph{initial} states, $T \subset Q$ is the set of 
\emph{terminal} states and $E \subset Q \times A \times Q$ is the set of 
labelled \emph{edges}.
If $(p,a,q) \in E$, we write $p \stackrel{a}{\to} q$.
The automaton is \emph{finite} if $Q$ is finite.
A subset $H$ of $A^*$ is said to be \emph{recognizable by a finite automaton} 
if there exists a finite automaton $\mathcal A$ such that $H$ is equal to the 
set of labels of paths starting in an initial state and ending in a terminal 
state.

A \emph{transducer} is an automaton $\mathcal{T}=(Q,A^* \times A'^*,E,I,T)$
where the edges of $E$ are labelled by couples of words in $A^* \times A'^*$. 
It is said to be \emph{finite} if the set $Q$ of states and the set $E$ of 
edges are finite. 
If $(p,(u,v),q) \in E$, we write $p \stackrel{u|v}{\longrightarrow} q$.
In this paper we consider \emph{letter-to-letter} transducers, where the edges 
are labelled by elements of $A \times A'$.
The \emph{input automaton} of such a transducer is obtained by taking the 
projection of edges on the first component.

\section{General case}
In this section, our aim is to prove that one can construct a finite automaton recognizing the set of $\beta$-expansions of minimal weight when $\beta$ is a Pisot number.

We need first some combinatorial results for bases $\beta$ satisfying ($\mathrm D_B$).
Note that $\beta$ is not assumed to be a Pisot number here.

\begin{proposition}\label{pARS}
If $\beta$ satisfies {\rm($\mathrm D_B$)} with some integer $B\ge2$, then for every word $x\in\mathbb Z^*$ there exists some $y\in\{1-B,\ldots,B-1\}^*$ with $x\sim_\beta y$ and $\|y\|\le\|x\|$.
\end{proposition}

\begin{corollary}\label{cARS}
If $\beta$ satisfies {\rm($\mathrm D_B$)} with some integer $B\ge2$, then for every word $x\in\mathbb Z^*$ there exists a $\beta$-expansion of minimal weight $y\in\{1-B,\ldots,B-1\}^*$ with $x\sim_\beta y$.
\end{corollary}

\begin{remark}
If $\beta$ satisfies {\rm($\mathrm D_B$)} for some positive integer $B$, then it is easy to see that $\beta$ satisfies {\rm($\mathrm D_C$)} for every integer $C>B$.
\end{remark}

\begin{remark}\label{rB}
If $\beta$ does not satisfy {\rm($\mathrm D_B$)}, then all words $x\in\{1-B,\ldots,B-1\}^*$ with $x\sim_\beta B$ are $\beta$-heavier than $B$. 
It follows that the set of $\beta$-expansions of minimal weight $x\sim_\beta B$ is $0^*B0^*$.
\end{remark}

\begin{proof}[Proof of Proposition~\ref{pARS}]
Let $A=\{1-B,\ldots,B-1\}$. 
If $x=x_1x_2\cdots x_n\in A^*$, then there is nothing to do. 
Otherwise, we use ($\mathrm D_B$): there exists some word $b=b_{-k}\cdots b_d\in A^*$ such that $b_{-k}\cdots b_{-1}(b_0-B)b_1\cdots b_d\sim_\beta 0$ and $\|b\|\le B$.
We use this relation to decrease the absolute value of a digit $x_h\not\in A$ without increasing the weight of~$x$, and we show that we eventually obtain a word in $A^*$ if we always choose the rightmost such digit.
More precisely, set $x_j^{(0)}=x_j$ for $1\le j\le n$, $x_j^{(0)}=0$ for $j\le0$ and $j>n$, $b_j=0$ for $j<-k$ and $j>d$.
Define, recursively for $i\ge0$, $h_i=\max\{j\in\mathbb Z:|x_j^{(i)}|\ge B\}$,
$$
x_{h_i}^{(i+1)}=x_{h_i}^{(i)}+\mathrm{sgn}(x_{h_i}^{(i)})(b_0-B),\
x_{h_i+j}^{(i+1)}=x_{h_i+j}^{(i)}+\mathrm{sgn}(x_{h_i}^{(i)})b_j\mbox{ for }
j\ne 0,
$$ 
as long as $h_i$ exists.
Then we have $\sum\limits_{j\in\mathbb Z}|x_j^{(0)}|=\|x\|$,
$\sum\limits_{j\in\mathbb Z}x_j^{(i+1)}\beta^{-j}=
\sum\limits_{j\in\mathbb Z}x_j^{(i)}\beta^{-j}$ and
$$
\sum_{j\in\mathbb Z}|x_j^{(i+1)}|=
|x_{h_i}^{(i+1)}|+\sum_{j\ne 0}|x_{h_i+j}^{(i+1)}|\le 
|x_{h_i}^{(i)}|+|b_0|-B+\sum_{j\ne 0}(|x_{h_i+j}^{(i)}|+|b_j|)\le
\sum_{j\in\mathbb Z}|x_j^{(i)}|.
$$
If $h_i$ does not exist, then we have $|x_j^{(i)}|<B$ for all $j\in\mathbb Z$, 
and the sequence $(x_j^{(i)})_{j\in\mathbb Z}$ without the leading and trailing 
zeros is a word $y\in A^*$ with the desired properties.

Since $\|x\|$ is finite, we have
$\sum_{j\in\mathbb Z}|x_j^{(i+1)}|<\sum_{j\in\mathbb Z}|x_j^{(i)}|$ 
only for finitely many $i\ge0$. 
In particular, the algorithm terminates after at most $\|x\|-B+1$ steps if
$\|b\|<B$.\footnote{For the proof of Theorem~\ref{main}, it is sufficient to consider the case $\|b\|<B$. However, Corollary~\ref{cARS} is particularly interesting in the case $\|b\|=B$, and we use it in the following sections for $B=2$.}

If $\|b\|=B$ and 
$\sum_{j\in\mathbb Z}|x_j^{(i+1)}|=\sum_{j\in\mathbb Z}|x_j^{(i)}|$, then we 
have 
$$
\sum_{j=-\infty}^{h_i-1}|x_j^{(i+1)}|=
\sum_{j=-\infty}^{h_i-1}|x_j^{(i)}|+\sum_{j=1}^k|b_{-j}|\ \mbox{ and }
\ \sum_{j=h_i+1}^\infty|x_j^{(i+1)}|=\sum_{j=h_i+1}^\infty|x_j^{(i)}|
+\sum_{j=1}^d|b_j|.
$$
Assume that $h_i$ exists for all $i\ge0$.
If $(h_i)_{i\ge0}$ has a minimum, then there exists an increasing sequence of indices $(i_m)_{m\ge0}$ such that $h_{i_m}\le h_\ell$ for all $\ell>i_m$, $m\ge0$, thus
$$
\|x\|\ge\sum_{j=-\infty}^{h_{i_m}-1}|x_j^{(i_m+1)}|\ge\sum_{j=-\infty}^{h_{i_{m-1}}-1}|x_j^{(i_{m-1}+1)}|+\sum_{j=1}^k|b_{-j}|\ge\cdots\ge (m+1)\sum_{j=1}^k|b_{-j}|.
$$
If $\sum_{j=1}^k|b_{-j}|>0$, this is not possible since $\|x\|$ is finite.
Similarly, $(h_i)_{i\ge0}$ has no maximum if $\sum_{j=1}^d|b_j|>0$.
Since $x_j^{(i+1)}$ can differ from $x_j^{(i)}$ only for $h_i-k\le j\le h_i+d$, we have $h_{i+1}\le h_i+d$ for all $i\ge0$.
If $h_i<h_{i'}$, $i<i'$, then there is therefore a sequence $(i_m)_{0\le m\le M}$, $i\le i_0<i_1<\cdots<i_M=i'$, with $M\ge(h_{i'}-h_i)/d$ such that $h_{i_m}\le h_\ell$ for all $\ell\in\{i_m,i_m+1,\ldots,i'\}$, $m\in\{0,\ldots,M\}$.
As above, we obtain $\|x\|\ge(M+1)\sum_{j=1}^k|b_{-j}|$, but $M$ can be arbitrarily large if $(h_i)_{i\ge0}$ has neither minimum nor maximum.
Hence we have shown that $h_i$ cannot exist for all $i\ge0$ if $\sum_{j=1}^k|b_{-j}|>0$ and $\sum_{j=1}^d|b_j|>0$.

It remains to consider the case $\|b\|=B$ with $k=0$ or $d=0$.
Assume, w.l.o.g., $d=0$.
Then we have $h_{i+1}\le h_i$.
If $h_i$ exists for all $i\ge0$, then both $\sum_{j=0}^k|x_{h_i-j}^{(i)}|$ and
$\sum_{j=1}^\infty|x_{h_i+j}^{(i)}|$ are eventually constant.
Therefore we must have some $i,i'$ with $h_{i'}<h_i$ such that 
$x_{h_{i'}-k}^{(i')}\cdots x_{h_{i'}}^{(i')}=x_{h_i-k}^{(i)}\cdots 
x_{h_i}^{(i)}$, $x_{h_{i'}+1}^{(i')}x_{h_{i'}+2}^{(i')}\cdots=
0^{h_i-h_{i'}}x_{h_i+1}^{(i)}x_{h_i+2}^{(i)}\cdots$, and $x_{h_i-j}^{(i)}=x_{h_{i'}-j}^{(i')}=0$ for all $j>k$.
This implies $x_{h_i-k}^{(i)}\cdots 
x_{h_i}^{(i)}\sim_\beta0$ or $\beta^{h_i-h_{i'}}=1$.
In the first case, $x_{h_i+1}^{(i)}x_{h_i+2}^{(i)}\cdots$ without the trailing zeros is a word $y\in A^*$ with the desired properties.
In the latter case, each $x\in\mathbb Z^*$ can be easily transformed into some $y\in\{-1,0,1\}^*$ with $y\sim_\beta x$ and $\|y\|=\|x\|$, and the proposition is proved.
\end{proof}

The following proposition shows slightly more than the existence of a positive integer $B$ such that $\beta$ satisfies {\rm($\mathrm D_B$)} when $\beta$ is a Pisot number.

\begin{proposition}\label{existsB}
For every Pisot number $\beta$, there exists some positive integer $B$ and some word $b\in\mathbb Z^*$ such that $B\sim_\beta b$ and $\|b\|<B$.
\end{proposition}

\begin{proof}
If $\beta$ is an integer, then we can choose $B=\beta$ and $b=1$. 
So let $\beta$ be a Pisot number of degree $d\ge2$, i.e., $\beta$ has $d-1$ Galois conjugates $\beta^{(j)}$, $2\le j\le d$, with $|\beta^{(j)}|<1$.  
For every $z\in\mathbb Q(\beta)$ set $z^{(j)}=P(\beta^{(j)})$ if $z=P(\beta)$, $P\in\mathbb Q[X]$. 

Let $B$ be a positive integer, $L=\lceil\log B/\log\beta\rceil$, and $x_1x_2\cdots$ the greedy $\beta$-expansion of $z=\beta^{-L}B\in[0,1)$.
Since 
$$
\tau_\beta^k(z)=\beta\tau^{k-1}(z)-x_k=\cdots=\beta^kz-\sum_{\ell=1}^kx_\ell\beta^{k-\ell},
$$ 
we have 
$$
\Big|(\tau_\beta^k(z))^{(j)}\Big|=\bigg|(\beta^{(j)})^k z^{(j)}-\sum_{\ell=1}^k x_\ell(\beta^{(j)})^{k-\ell}\bigg|<\big|\beta^{(j)}\big|^k\big|z^{(j)}\big|+\frac{\lfloor\beta\rfloor}{1-|\beta^{(j)}|}
$$
for all $k\ge0$ and $2\le j\le d$. 
Set $k=\max_{2\le j\le d}\lceil-\log|z^{(j)}|/\log|\beta^{(j)}|\rceil$.
Then $\tau_\beta^k(z)$ is an element of the finite set
$$
Y=\bigg\{y\in\mathbb Z[\beta^{-1}]\cap[0,1):\ |y^{(j)}|<1+\frac{\lfloor\beta\rfloor}{1-|\beta^{(j)}|}\ \mbox{ for }\ 2\le j\le d\bigg\}.
$$
For every $y\in Y$, we can choose a $\beta$-expansion $y=\decdot a_1\cdots a_m$.
Let $W$ be the maximal weight of all these expansions and  $\tau_\beta^k(z)=\decdot a_1'\cdots a_m'$.
Since $z=\decdot x_1\dots x_k+\tau_\beta^k(z)$, the digitwise addition of $x_1\cdots x_k$ and $a_1'\cdots a_m'$ provides a word $b$ with $b\sim_\beta B$ and 
$$
\|b\|\le k\lfloor\beta\rfloor+W=\max_{2\le j\le d}\bigg\lceil\bigg\lceil\frac{\log B}{\log\beta}\bigg\rceil-\frac{\log B}{\log|\beta^{(j)}|}\bigg\rceil\lfloor\beta\rfloor+W=\mathcal O(\log B).
$$
If $B$ is sufficiently large, we have therefore $\|b\|<B$.
\end{proof}

In order to understand the relation $\sim_\beta$ on $A^*$, $A=\{1-B,\ldots,B-1\}$, we have to consider the words $z\in(A-A)^*$ with $z\sim_\beta0$.
Therefore we set
$$
Z_\beta=\Big\{z_1\cdots z_n\in\{2(1-B),\ldots,2(B-1)\}^*\ \Big|\ n\ge0,\;
\sum_{j=1}^n z_j\beta^{-j}=0\Big\}
$$ 
and recall a result from~\cite{Frougny92}.
All the automata considered in this paper process words from left to 
right, that is to say, most significant digit first.

\begin{lemma}[\cite{Frougny92}]\label{threc}
If $\beta$ is a Pisot number, then $Z_\beta$ is recognized by a finite automaton.
\end{lemma}

For convenience, we quickly explain the construction of the automaton $\mathcal A_\beta$ recognizing $Z_\beta$.
The states of $\mathcal A_\beta$ are $0$ and all $s\in\mathbb Z[\beta]\cap(\frac{2(1-B)}{\beta-1},\frac{2(B-1)}{\beta-1})$ which are 
accessible from $0$ by paths consisting of transitions $s\stackrel e\to s'$
with $e\in A-A$ such that $s'=\beta s+e$. 
The state $0$ is both initial and terminal. 
When $\beta$ is a Pisot number, then the set of states is finite.
Note that $\mathcal A_\beta$ is symmetric, meaning that if $s\stackrel e\to s'$ is a transition, then $-s\stackrel{-e}\to -s'$ is also a transition.
The automaton $\mathcal A_\beta$ is accessible and co-accessible.

The {\em redundancy automaton} (or transducer) $\mathcal R_\beta$ is similar to $\mathcal A_\beta$.
Each transition $s\stackrel e\to s'$ of $\mathcal A_\beta$ is replaced in $\mathcal R_\beta$ by a set of transitions $s\stackrel{a|b}\longrightarrow s'$, with $a,b\in A$ and $a-b=e$. 
{}From Lemma~\ref{threc}, one obtains the following lemma.

\begin{lemma}
The redundancy transducer $\mathcal R_\beta$ recognizes the set
$$
\big\{(x_1\cdots x_n,y_1\cdots y_n) \in A^* \times A^*\ \big|\ n\ge0,\;
\decdot x_1\cdots x_n=\decdot y_1\cdots y_n\big\}.
$$
If $\beta$ is a Pisot number, then $\mathcal R_\beta$ is finite.
\end{lemma}

{}From the redundancy transducer $\mathcal R_\beta$, one constructs another transducer $\mathcal T_\beta$ with states of the form $(s,\delta)$, where $s$ is a state of $\mathcal R_\beta$ and $\delta\in\mathbb Z$. 
The transitions are of the form $(s,\delta)\stackrel{a|b}\longrightarrow(s',\delta')$ if 
$s\stackrel{a|b}\longrightarrow s'$ is a transition in $\mathcal R_\beta$ and $\delta'=\delta+|b|-|a|$. 
The initial state is $(0,0)$, and terminal states are of the form $(0,\delta)$ with $\delta<0$.

\begin{lemma}
The transducer $\mathcal T_\beta$ recognizes the set
$$
\big\{(x_1\cdots x_n,y_1 \cdots y_n) \in  A^* \times A^*\ \big|\
\decdot x_1\cdots x_n=\decdot y_1\cdots y_n,\
\|y_1\cdots y_n\|<\|x_1\cdots x_n\|\big\}.
$$
\end{lemma}

Of course, the transducer $\mathcal T_\beta$ is not finite, and the core of the proof of the main result consists in showing that we need only a finite part of $\mathcal T_\beta$. 

We also need the following well-known lemma, and give a proof for it because the construction in the proof will be used in the following sections.

\begin{lemma}\label{rat}
Let $H\subset A^*$ and $M=A^*\setminus A^*HA^*$.
If $H$ is recognized by a finite automaton, then so is $M$.
\end{lemma}

\begin{proof}
Suppose that $H$ is recognized by a finite automaton $\mathcal H$.
Let $P$ be the set of strict prefixes of $H$. 
We construct the minimal automaton $\mathcal M$ of $M$ as follows. 
The set of states of $\mathcal M$ is the quotient $P/_\equiv$ where $p\equiv q$ if the paths labelled by $p$ end in the same set of states in $\mathcal{H}$ as the paths labelled by $q$.
Since $\mathcal{H}$ is finite, $P/_\equiv$ is finite.
Transitions are defined as follows. 
Let $a$ be in $A$. 
If $pa$ is in $P$, then there is a transition $[p]_\equiv\stackrel a\to[pa]_\equiv$.
If $pa$ is not in $H\cup P$, then there is a transition $[p]_\equiv\stackrel a\to[v]_\equiv$ with $v$ in $P$ maximal in length such that $pa=uv$.
Every state is terminal.
\end{proof}

Now, we can prove the following theorem.
The main result, Theorem~\ref{main}, will be a special case of it.

\begin{theorem}\label{D}
Let $\beta$ be a Pisot number and $B$ a positive integer such that {\rm($\mathrm D_B$)} holds.
Then one can construct a finite automaton recognizing the set of $\beta$-expansions of minimal weight in $\{1-B,\ldots,B-1\}^*$.
\end{theorem}

\begin{proof}
Let $A=\{1-B,\ldots,B-1\}$, $x\in A^*$ be a strictly $\beta$-heavy word and $y\in A^*$ be a $\beta$-expansion of minimal weight with $x\sim_\beta y$.
Such a $y$ exists because of Proposition~\ref{pARS}.
Extend $x,y$ to words $x',y'$ by adding leading and trailing zeros such that $x'=x_1\cdots x_n$, $y'=y_1\cdots y_n$ and $\decdot x_1\cdots x_n=\decdot y_1\cdots y_n$.
Then there is a path in the transducer $\mathcal T_\beta$ composed of transitions $(s_{j-1},\delta_{j-1})\stackrel{x_j|y_j}\longrightarrow(s_j,\delta_j)$, $1\le j\le n$, with $s_0=0$, $\delta_0=0$, $s_n=0$, $\delta_n<0$.

We determine bounds for $\delta_j$, $1\le j\le n$, which depend only on the state $s=s_j$.
Choose a $\beta$-expansion of $s$, $s=a_1\cdots a_i\decdot a_{i+1}\cdots a_m$, and set $w_s=\|a_1\cdots a_m\|$.
If $\delta_j>w_s$, then we have $\|y_1\cdots y_j\|>\|x_1\cdots x_j\|+w_s$.
Since $s_j=(x_1-y_1)\cdots(x_j-y_j)\decdot$, the digitwise subtraction of $0^{\max(i-j,0)}x_1\cdots x_j0^{m-i}$ and $0^{\max(j-i,0)}a_1\cdots a_m$ provides a word which is $\beta$-lighter than $y_1\cdots y_j$, which contradicts the assumption that $y$ is a $\beta$-expansion of minimal weight.

Let $W=\max\{w_s\mid s\mbox{ is a state in }\mathcal A_\beta\}$. 
If $\delta_j\le-W-B$, then let $h\le j$ be such that $x_h\ne 0$, $x_i=0$ for $h<i\le j$.
Since $|x_h|<B$, we have $\delta_{h-1}<\delta_j+B\le-W\le-w_{s_{h-1}}$, hence $\|x_1\cdots x_{h-1}\|>\|y_1\cdots y_{h-1}\|+w_{s_{h-1}}$.
Let $a_1\cdots a_m$ be the word which was used for the definition of $w_{s_{h-1}}$, i.e., $s_{h-1}=a_1\cdots a_i\decdot a_{i+1}\cdots a_m$, $w_{s_{h-1}}=\|a_1\cdots a_m\|$. 
Then the digitwise addition of $0^{\max(i-h+1,0)}y_1\cdots y_{h-1}0^{m-i}$ and $0^{\max(h-1-i,0)}a_1\cdots a_m$ provides a word which is $\beta$-lighter than $x_1\cdots x_{h-1}$.
Since $x_h\ne 0$, this contradicts the assumption that $x$ is strictly $\beta$-heavy.

Let $\mathcal S_\beta$ be the restriction of $\mathcal T_\beta$ to the states $(s,\delta)$ with $-W-B<\delta\le w_s$ with some additional initial and terminal states: 
Every state which can be reached from $(0,0)$ by a path with input in $0^*$ is initial, and every state with a path to $(0,\delta)$, $\delta<0$, with an input in $0^*$ is terminal.
Then the set $H$ which is recognized by the input automaton of $\mathcal S_\beta$ consists only of $\beta$-heavy words and contains all strictly $\beta$-heavy words in $A^*$. 
Therefore $M=A^* \setminus A^*HA^*$ is the set of $\beta$-expansions of minimal weight in $A^*$, and $M$ is recognizable by a finite automaton by Lemma~\ref{rat}.
\end{proof}

\begin{theorem}\label{main}
Let $\beta$ be a Pisot number.
Then one can construct a finite automaton recognizing the set of $\beta$-expansions of minimal weight.
\end{theorem}

\begin{proof}
Proposition~\ref{existsB} shows that $\beta$ satisfies ($\mathrm D_B$) for some positive integer $B$, and that no $\beta$-expansion of minimal weight $y\in\mathbb Z^*$ can contain a digit $y_j$ with $|y_j|\ge B$, since we obtain a $\beta$-lighter word if we replace $B$ by $b$ as in the proof of Proposition~\ref{pARS}.
Therefore Theorem~\ref{D} implies Theorem~\ref{main}.
\end{proof}

\section{Golden Ratio case}\label{GR}

In this section we give explicit constructions for the case where 
$\beta$ is the Golden Ratio $\frac{1+\sqrt5}2$. 
We have $1=\decdot 11$, hence $2=10\decdot 01$ and $\beta$ satisfies ($\mathrm D_2$), see also Example~\ref{ex1}.
Corollary~\ref{cARS} shows that every $z\in\mathbb Z[\beta^{-1}]$ can be represented by a $\beta$-expansion of minimal weight in $\{-1,0,1\}^*$. 
For most applications, only these expansions are interesting.
Remark that the digits of arbitrary $\beta$-expansions of minimal weight are in $\{-2,-1,0,1,2\}$ by the proof of Theorem~\ref{main}, since $3=100\decdot01$.

For typographical reasons, we write the digit $-1$ as $\bar1$ in words and transitions.

\subsection{$\beta$-expansions of minimal weight for $\beta=\frac{1+\sqrt5}2$}

Our aim in this section is to construct explicitly the finite automaton recognizing the $\beta$-expansions of minimal weight in $A^*$, $A=\{-1,0,1\}$. 

\begin{figure}[ht]
\centerline{\includegraphics[scale=.94]{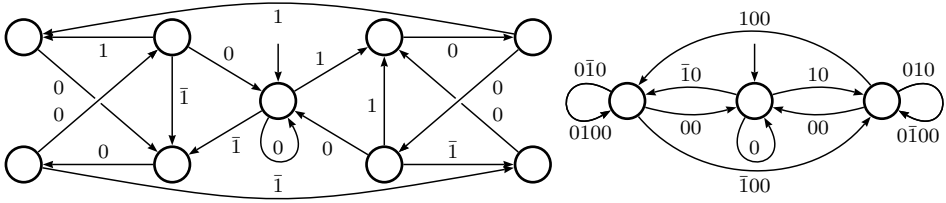}}
\caption{Automaton $\mathcal M_\beta$ recognizing $\beta$-expansions of minimal 
weight for $\beta=\frac{1+\sqrt5}2$ (left) and a compact representation of
$\mathcal M_\beta$ (right).} 
\label{figexp}
\end{figure}

\begin{theorem}\label{minimal}
If $\beta=\frac{1+\sqrt5}2$, then the set of $\beta$-expansions of minimal 
weight in $\{-1,0,1\}^*$ is recognized by the finite automaton 
$\mathcal M_\beta$ of Figure~\ref{figexp} where all states are terminal.
\end{theorem}

\begin{figure}[ht]
\centerline{\includegraphics{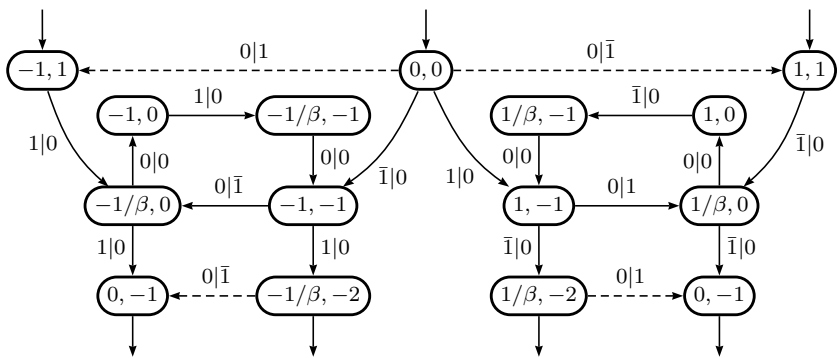}}
\caption{Transducer with strictly $\beta$-heavy words as inputs,
$\beta=\frac{1+\sqrt5}2$.} \label{figforb}
\end{figure}

It is of course possible to follow the proof of Theorem~\ref{D}, but the states of $\mathcal A_\beta$ are
$$
0,\ \pm\frac1{\beta^3},\ \pm\frac1{\beta^2},\ \pm\frac1\beta,\ \pm1,\ \pm\beta,\ \pm\beta^2,\ \pm\beta\pm\frac1{\beta^2},\ \pm\beta\pm\frac1{\beta^3},\ \pm\beta^2\pm\frac1{\beta^2},\ \pm\beta^2\pm\frac1{\beta^3},
$$
thus $W=2$ and the transducer $\mathcal S_\beta$ has 160 states.
For other bases $\beta$, the number of states can be much larger. 
Therefore we have to refine the techniques if we do not want computer-assisted proofs.
It is possible to show that a large part of $\mathcal S_\beta$ is not needed, e.g. by excluding some $\beta$-heavy factors such as $11$ from the output, and to obtain finally the transducer in Figure~\ref{figforb}.
However, it is easier to prove Theorem~\ref{minimal} by an indirect strategy, which includes some results which are interesting by themselves. 

\begin{lemma}\label{heavy}
All words in $\{-1,0,1\}^*$ which are not recognized by the automaton 
$\mathcal M_\beta$ in Figure~\ref{figexp} are $\beta$-heavy.
\end{lemma}

\begin{proof}
The transducer in Figure~\ref{figforb} is a part of the transducer $\mathcal S_\beta$ in the proof of Theorem~\ref{D}.
This means that every word which is the input of a path (with full or dashed transitions) going from $(0,0)$ to $(0,-1)$ is $\beta$-heavy, because the output has the same value but less weight.
Since a $\beta$-heavy word remains $\beta$-heavy if we omit the leading and trailing zeros, the dashed transitions can be omitted.
Then the set of inputs is
\begin{align*}
H & = 1(0100)^*1\ \cup\ 1(0100)^*0101\ \cup\ 1(00\bar10)^*\bar1\ \cup\
1(00\bar10)^*0\bar1 \\ & \ \cup\ \bar1(0\bar100)^*\bar1\ \cup\ 
\bar1(0\bar100)^*0\bar10\bar1\ \cup\ \bar1(0010)^*1\ \cup\ \bar1(0010)^*01
\end{align*}
and $\mathcal M_\beta$ is constructed as in the proof of Lemma~\ref{rat}.
\end{proof}

Similarly to the NAF in base~2, where the expansions of minimal weight avoid the set $\{11,1\bar1,\bar1\bar1,\bar11\}$, we show in the next result that, for $\beta=\frac{1+\sqrt5}2$, every real number admits a $\beta$-expansion which avoids a certain finite set $X$.

\begin{proposition}\label{particular}
If $\beta=\frac{1+\sqrt5}2$, then every $z\in\mathbb R$ has a $\beta$-expansion 
of the form $z=y_1\cdots y_k\decdot y_{k+1}y_{k+2}\cdots$ with 
$y_j\in\{-1,0,1\}$ such that $y_1y_2\cdots$ avoids the set 
$X= \{11, 101, 1001, 1\bar1, 10\bar1,\mbox{ and their opposites}\}$.
If $z\in\mathbb Z[\beta]=\mathbb Z[\beta^{-1}]$, then this expansion is unique 
up to leading zeros.
\end{proposition}

\begin{proof}
We determine this $\beta$-expansion similarly to the greedy $\beta$-expansion 
in the Introduction.
Note that the maximal value of $\decdot x_1x_2\cdots$ for a sequence $x_1x_2\cdots$ avoiding the elements of $X$ is $\decdot(1000)^\omega=\beta^2/(\beta^2+1)$.
If we define the transformation 
$$
\tau:\ \left[\frac{-\beta^2}{\beta^2+1},\frac{\beta^2}{\beta^2+1}\right)\to
\left[\frac{-\beta^2}{\beta^2+1},\frac{\beta^2}{\beta^2+1}\right),\quad
\tau(z)=\beta z-\left\lfloor\frac{\beta^2+1}{2\beta}z+1/2\right\rfloor,
$$
and set $y_j=\big\lfloor\frac{\beta^2+1}{2\beta}\tau^{j-1}(z)+1/2\big\rfloor$ 
for $z\in\left[\frac{-\beta^2}{\beta^2+1},\frac{\beta^2}{\beta^2+1}\right)$, $j\ge1$, then $z=\decdot y_1y_2\cdots$.
If $y_j=1$ for some $j\ge1$, then we have $\tau^j(z)\in\beta\times
\big[\frac\beta{\beta^2+1},\frac{\beta^2}{\beta^2+1}\big)-1=
\big[\frac{-1}{\beta^2+1},\frac{1/\beta}{\beta^2+1}\big)$, hence $y_{j+1}=0$, 
$y_{j+2}=0$, and 
$\tau^{j+2}(z)\in\big[\frac{-\beta^2}{\beta^2+1},\frac\beta{\beta^2+1}\big)$,
hence $y_{j+3}\in\{\bar1,0\}$.
This shows that the given factors are avoided.
A similar argument for $y_j=-1$ shows that the opposites are avoided as 
well, hence we have shown the existence of the expansion for
$z\in\left[\frac{-\beta^2}{\beta^2+1},\frac{\beta^2}{\beta^2+1}\right)$. 
For arbitrary $z\in\mathbb R$, the expansion is given by shifting the expansion
of $\beta^{-k}z$, $k\ge0$, to the left.

If we choose $y_j=0$ in case 
$\tau^{j-1}(z)>\beta/(\beta^2+1)=\decdot(0100)^\omega$, then it is impossible 
to avoid the factors $11$, $101$ and $1001$ in the following. 
If we choose $y_j=1$ in case $\tau^{j-1}(z)<\beta/(\beta^2+1)$, then 
$\beta\tau^{j-1}(z)-1<-1/(\beta^2+1)=\decdot(00\bar10)^\omega$, and thus
it is impossible to avoid the factors $1\bar1$, $10\bar1$, $\bar1\bar1$, 
$\bar10\bar1$ and $\bar100\bar1$.
Since $\beta/(\beta^2+1)\not\in\mathbb Z[\beta]$, we have 
$\tau^{j-1}(z)\ne\beta/(\beta^2+1)$ for $z\in\mathbb Z[\beta]$.
Similar relations hold for the opposites, thus the expansion is unique.
\end{proof}

\begin{remark}
Similarly, the transformation $\tau(z)=\beta z-\lfloor z+1/2\rfloor$ on 
$[-\beta/2,\beta/2)$ provides for every $z\in\mathbb Z[\beta]$ a unique 
expansion avoiding the factors $11$, $101$, $1\bar1$, $10\bar1$, $100\bar1$ and their opposites.
\end{remark}

\begin{figure}[ht]
\centerline{\includegraphics[scale=.9]{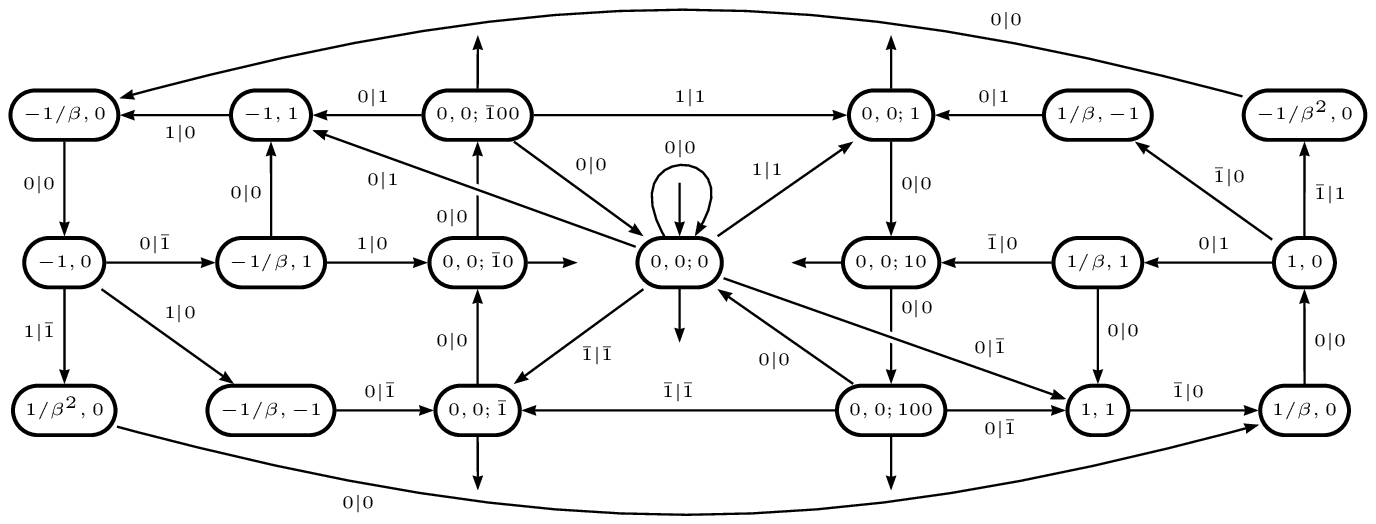}}
\caption{Transducer $\mathcal N_\beta$ normalizing $\beta$-expansions of 
minimal weight, $\beta=\frac{1+\sqrt5}2$.}
\label{figtrans}
\end{figure}

\begin{proposition}\label{conversion}
If $x$ is accepted by $\mathcal M_\beta$, then there exists $y\in\{-1,0,1\}^*$ avoiding $X= \{11, 101, 1001, 1\bar1, 10\bar1\mbox{ and their opposites}\}$ with $x\sim_\beta y$ and $\|x\|=\|y\|$.
The transducer $\mathcal N_\beta$ in Figure~\ref{figtrans} realizes the conversion from $0x0$ to $y$.
\end{proposition}

\begin{proof}
Set $Q_0=\{(0,0;0),\,(-1,1),\,(1,1)\}=Q_0'$,
\begin{align*}
Q_1&=\{(0,0;1),\,(-1/\beta,0)\},&Q_1'&=\{(0,0;\bar10)\}, \\
Q_{10}&=\{(0,0;10),\,(-1,0)\},&Q_{10}'&=\{(0,0;\bar100)\}, \\
Q_{100}&=\{(0,0;100),\,(-1/\beta,1)\},&Q_{100}'&=\{(0,0;0),\,(-1,1)\}, \\
Q_{101}&=\{(-1/\beta,-1),\,(1/\beta^2,0)\},&Q_{101}'&=\{(0,0;1)\},
\end{align*}
and, symmetrically, $Q_{\bar1}=\{(0,0;\bar1),\,(1/\beta,0)\}$, $Q_{\bar1}'=\{0,0;10\}$, \dots\,.
Then the paths in $\mathcal N_\beta$ with input in $00^*$ lead to the three 
states in $Q_0$, the paths with input $01$ lead to the two states in $Q_1$, and 
more generally the paths in $\mathcal N_\beta$ with input $0x$ such that $x$ is
accepted by $\mathcal M_\beta$ lead to all states in $Q_u$ or to all states in 
$Q_u'$, where $u$ labels the shortest path in $\mathcal M_\beta$ leading to the 
state reached by $x$. 
Indeed, if $u\stackrel a\to v$ is a transition in $\mathcal M_\beta$, then we 
have $Q_u\stackrel a\to Q_v$ or $Q_u\stackrel a\to Q_v'$, and
$Q_u'\stackrel a\to Q_v$ or $Q_u'\stackrel a\to Q_v'$, where
$Q\stackrel a\to R$ means that for every $r\in R$ there exists a transition 
$q\stackrel{a|b}\longrightarrow r$ in $\mathcal N_\beta$ with $q\in Q$.

Since every $Q_u$ and every $Q_u'$ contains a state $q$ with a transition of 
the form $q\stackrel{0|b}\longrightarrow(0,0;w)$, there exists a path with 
input $0x0$ going from $(0,0;0)$ to $(0,0;w)$ for every word $x$ accepted by 
$\mathcal M_\beta$.
By construction, the output $y$ of this path satisfies $x\sim_\beta y$ and 
$\|x\|=\|y\|$.
It can be easily checked that all outputs of $\mathcal N_\beta$ avoid the 
factors in $X$.
\end{proof}

\begin{proof}[Proof of Theorem~\ref{minimal}]
For every $x\in\mathbb Z^*$, by Proposition~\ref{pARS} and Lemma~\ref{heavy},  there exists a $\beta$-expansion of minimal weight $y$ accepted by $\mathcal M_\beta$ with $y\sim_\beta x$.
By Proposition~\ref{conversion}, there also exists a $\beta$-expansion of minimal weight $y'\in\{-1,0,1\}^*$ avoiding $X$ with $y'\sim_\beta y\sim_\beta x$.
By Proposition~\ref{particular}, the output of $\mathcal N_\beta$ is the same (if we neglect leading and trailing zeros) for every input $0x'0$ such that $x'\sim_\beta x$ and $x'$ is accepted by~$\mathcal M_\beta$.
Therefore $\|x'\|=\|y'\|$ for all these $x'$, and the theorem is proved.
\end{proof}

\subsection{Branching transformation}

All $\beta$-expansions of minimal weight can be obtained by a branching
transformation.

\begin{theorem}
Let $x=x_1\cdots x_n\in\{-1,0,1\}^*$ and $z=\decdot x_1\cdots x_n$,
$\beta=\frac{1+\sqrt5}2$.
Then $x$ is a $\beta$-expansion of minimal weight if and only if 
$-\frac{2\beta}{\beta^2+1}<z<\frac{2\beta}{\beta^2+1}$ and 
$$
x_j=\left\{\begin{array}{cl}1 & \mbox{if }\frac2{\beta^2+1}<
\beta^{j-1}z-x_1\cdots x_{j-1}\decdot<\frac{2\beta}{\beta^2+1} \\
0\mbox{ or }1 & \mbox{if }\frac\beta{\beta^2+1}<
\beta^{j-1}z-x_1\cdots x_{j-1}\decdot<\frac2{\beta^2+1} \\
0 & \mbox{if }\frac{-\beta}{\beta^2+1}<
\beta^{j-1}z-x_1\cdots x_{j-1}\decdot<\frac\beta{\beta^2+1} \\
-1\mbox{ or }0 & \mbox{if }\frac{-2}{\beta^2+1}<
\beta^{j-1}z-x_1\cdots x_{j-1}\decdot<\frac{-\beta}{\beta^2+1} \\ 
-1 & \mbox{if }\frac{-2\beta}{\beta^2+1}<
\beta^{j-1}z-x_1\cdots x_{j-1}\decdot<\frac{-2}{\beta^2+1}
\end{array}\right. \mbox{ for all }j,\ 1\le j\le n. 
$$
\end{theorem}

\begin{figure}[ht]
\centerline{\includegraphics{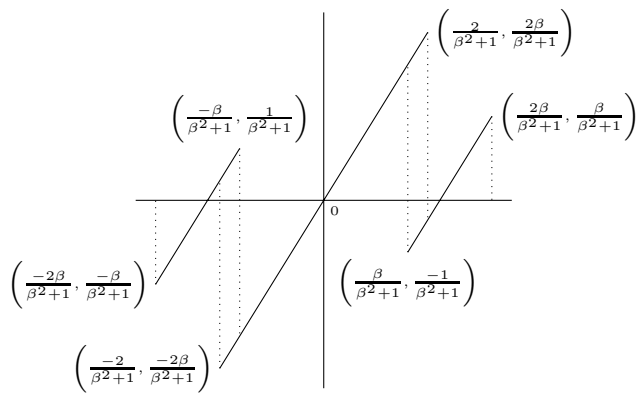}}
\caption{Branching transformation giving all $\frac{1+\sqrt5}2$-expansions of 
minimal weight.}
\label{figbranch}
\end{figure}

The sequence $(\beta^{j-1}z-x_1\cdots x_{j-1}\decdot)_{1\le j\le n}$ is a 
trajectory $(\tau^{j-1}(z))_{1\le j\le n}$, where the branching transformation 
$\tau:z\mapsto\beta z-x_1$ with $x_1\in\{-1,0,1\}$ is given in 
Figure~\ref{figbranch}.

\begin{proof}
To see that all words $x_1\cdots x_n$ given by the branching transformation are 
$\beta$-expansions of minimal weight, we have drawn in 
Figure~\ref{figintervals} an automaton where every state is labeled by the 
interval containing all numbers $\beta^jz-x_1\cdots x_j\decdot$ such 
that $x_1\cdots x_j$ labels a path leading to this state.
This automaton turns out to be the automaton $\mathcal M_\beta$ in 
Figure~\ref{figexp} (up to the labels of the states), which accepts exactly the 
$\beta$-expansions of minimal weight.
Recall that $\decdot(0010)^\omega=\frac1{\beta^2+1}$ and thus
$\decdot1(0100)^\omega=\frac{2\beta}{\beta^2+1}$.  

\begin{figure}[ht]
\centerline{\includegraphics{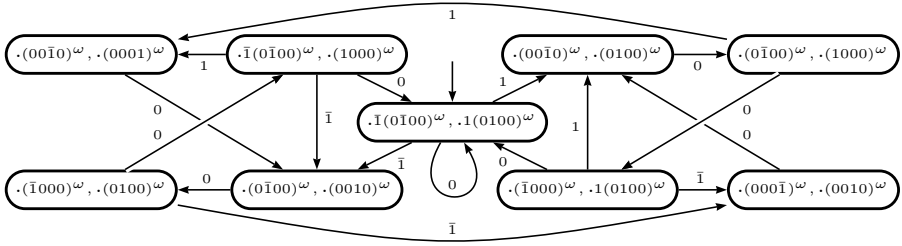}}
\caption{Automaton $\mathcal M_\beta$ with intervals as labels.}
\label{figintervals}
\end{figure}

If the conditions on $z$ and $x_j$ are not satisfied, then we have either 
$|\decdot x_j\cdots x_n|>\decdot1(0100)^\omega$, or $x_j=1$ and
$\decdot x_{j+1}\cdots x_n<\decdot(00\bar10)^\omega$, or $x_j=-1$ and
$\decdot x_{j+1}\cdots x_n>\decdot(0010)^\omega$ for some $j$, $1\le j\le n$.
In every case, it is easy to see that $x_j\cdots x_n$ must contain a factor in 
the set $H$ of the proof of Lemma~\ref{heavy}, hence $x_1\cdots x_n$ is 
$\beta$-heavy.
\end{proof}

\subsection{Fibonacci numeration system}

The reader is referred to~\cite[Chapter~7]{Lothaire02} for definitions on numeration systems defined by a sequence of integers.
Recall that the linear numeration system canonically associated with the Golden 
Ratio is the Fibonacci (or Zeckendorf) numeration system defined by the 
sequence of Fibonacci numbers $F=(F_n)_{n \ge 0}$ with 
$F_n=F_{n-1}+F_{n-2}$, $F_0=1$ and $F_1=2$.
Any non-negative integer $N<F_n$ can be represented as 
$N=\sum_{j=1}^n x_j F_{n-j}$ with the property that 
$x_1\cdots x_n\in\{0,1\}^*$ does not contain the factor $11$.
For words $x=x_1\cdots x_n\in\mathbb Z^*$, $y=y_1\cdots y_m\in\mathbb Z^*$, we 
define a relation 
$$
x\sim_F y\quad\mbox{ if and only if }\quad
\sum_{j=1}^n x_j F_{n-j}=\sum_{j=1}^m y_j F_{m-j}.
$$
The properties \emph{$F$-heavy} and \emph{$F$-expansion of minimal weight} are 
defined as for $\beta$-expansions, with $\sim_F$ instead of $\sim_\beta$.
An important difference between the notions $F$-heavy and $\beta$-heavy is that 
a word containing a $F$-heavy factor need not be $F$-heavy, e.g. $2$ is 
$F$-heavy since $2\sim_F 10$, but $20$ is not $F$-heavy.
However, $uxv$ is $F$-heavy if $x0^{\mathrm{length}(v)}$ is $F$-heavy.
Therefore we say that $x\in\mathbb Z^*$ is \emph{strongly $F$-heavy} if every 
element in $x0^*$ is $F$-heavy.
Hence every word containing a strongly $F$-heavy factor is $F$-heavy.

The Golden Ratio satisfies ($\mathrm D_2$) since $2=10\decdot01$. 
For the Fibonacci numbers, the corresponding relation is 
$2F_n=F_{n+1}+F_{n-2}$, hence $20^n\sim_F 10010^{n-2}$ for all $n\ge2$. 
Since $20\sim_F 101$ and $2\sim_F 10$, we obtain similarly to the proof of
Proposition~\ref{pARS} that for every $x\in\mathbb Z^*$ there exists some 
$y\in\{-1,0,1\}^*$ with $x\sim_F y$ and $\|y\|\le\|x\|$.
We will show the following theorem. 

\begin{theorem}\label{equal}
The set of $F$-expansions of minimal weight in $\{-1,0,1\}^*$ is equal to the 
set of $\beta$-expansions of minimal weight in $\{-1,0,1\}^*$ for 
$\beta=\frac{\sqrt5+1}2$.
\end{theorem}

The proof of this theorem runs along the same lines as the proof of 
Theorem~\ref{minimal}.
We use the unique expansion of integers given by Proposition~\ref{Heuberger}
(due to Heuberger~\cite{Heuberger04}) and provide an alternative proof of 
Heuberger's result that these expansions are $F$-expansions of minimal weight.

\begin{proposition}[\cite{Heuberger04}]\label{Heuberger}
Every $N\in\mathbb Z$ has a unique representation $N=\sum_{j=1}^n y_j F_{n-j}$ 
with $y_1\ne0$ and $y_1\cdots y_n\in\{-1,0,1\}^*$ avoiding 
$X= \{11, 101, 1001, 1\bar1, 10\bar1$, and their opposites$\}$.
\end{proposition}

\begin{proof}
Let $g_n$ be the smallest positive integer with an $F$-expansion of length $n$ 
starting with $1$ and avoiding $X$, and $G_n$ be the largest integer of this 
kind. 
Since $g_{n+1}\sim_F 1(00\bar10)^{n/4}$, $G_n\sim_F (1000)^{n/4}$ and
$1(\bar10\bar10)^{n/4}\sim_F 1$, we obtain $g_{n+1}-G_n=1$.
(A fractional power $(y_1\cdots y_k)^{j/k}$ denotes the word
$(y_1\cdots y_k)^{\lfloor j/k\rfloor}y_1\cdots 
y_{j-\lfloor j/k\rfloor k}$.) 
Therefore the length $n$ of an expansion $y_1y_2\cdots y_n$ of $N\ne0$ 
with $y_1\ne0$ avoiding $X$ is determined by 
$G_{n-1}<|N|\le G_n$. 
Since $g_n-F_{n-1}=-G_{n-3}$ and $G_n-F_{n-1}=G_{n-4}$, we have
$-G_{n-3}\le N-F_{n-1}\le G_{n-4}$ if $y_1=1$, hence $y_2=y_3=0$, 
$y_4\ne 1$, and we obtain recursively that $N$ has a unique expansion 
avoiding $X$. 
\end{proof}

\begin{figure}[ht]
\centerline{\includegraphics[scale=.9]{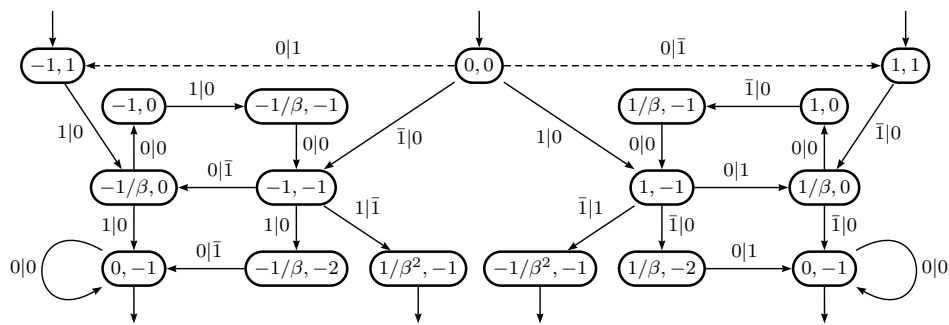}}
\caption{All inputs of this transducer are strongly $F$-heavy.}\label{figforbF}
\end{figure}

\begin{proof}[Proof of Theorem~\ref{equal}]
Let $a_1\cdots a_n\in\mathbb Z^*$, $z=\sum_{j=1}^n a_j\beta^{n-j}$, 
$N=\sum_{j=1}^n a_jF_{n-j}$.
By using the equations $\beta^k=\beta^{k-1}+\beta^{k-2}$ and 
$F_k=F_{k-1}+F_{k-2}$, we obtain integers $m_0$ and $m_1$ such that
$z=m_1\beta+m_0$ and $N=m_1F_1+m_0F_0=2m_1+m_0$.
Clearly, $z=0$ implies $m_1=m_0=0$ and thus $N=0$, but the converse is 
not true: $N=0$ only implies $m_0=-2m_1$, i.e., $z=-m_1/\beta^2$.
Therefore we have $x_1\cdots x_n\sim_F y_1\cdots y_n$ if and only if 
$(x_1-y_1)\cdots (x_n-y_n)\decdot=m/\beta^2$ for some $m\in\mathbb Z$, hence
the redundancy transducer $\mathcal R_F$ for the Fibonacci numeration system 
is similar to $\mathcal R_\beta$, except that all states $m/\beta^2$,
$m\in\mathbb Z$, are terminal.

The transducer in Figure~\ref{figforbF} shows that all strictly $\beta$-heavy
words in $\{-1,0,1\}^*$ are strongly $F$-heavy.
Therefore all words which are not accepted by $\mathcal M_\beta$ are 
$F$-heavy.
Let $\mathcal N_F$ be as $\mathcal N_\beta$, except that the states
$(\pm1/\beta^2,0)$ are terminal. 
Every set $Q_u$ and $Q_u'$ contains a state of the form $(0,0;w)$ or
$(\pm1/\beta^2,0)$.
If $x$ is accepted by $\mathcal N_\beta$, then $\mathcal N_F$ transforms 
therefore $0x$ into a word $y$ avoiding the factors given in 
Proposition~\ref{Heuberger}.
Hence $x$ is an $F$-expansion of minimal weight.
\end{proof}

\begin{remark}
If we consider only expansions avoiding the factors $11$, $101$, $1\bar1$, 
$10\bar1$, $100\bar1$, then the difference between the largest integer with 
expansion of length $n$ and the smallest positive integer with expansion of 
length $n+1$ is $2$ if $n$ is a positive multiple of $3$.
Therefore there exist integers without an expansion of this kind, e.g. $N=4$.
However, a small modification provides another ``nice'' set of $F$-expansions 
of minimal weight:
Every integer has a unique representation of the form 
$N=\sum_{j=1}^n y_jF_{n-j}$ with $y_1\ne 0$,
$y_1\cdots y_n\in\{\bar1,0,1\}^*$ avoiding the factors 
$11,\bar1\bar1,\bar10\bar1,1\bar1,\bar11,10\bar1,\bar101,100\bar1$ and
$y_{j-2}y_{j-1}y_j=101$ or $y_{j-3}\cdots y_j=\bar1001$ only if $j=n$.
\end{remark}

\subsection{Weight of the expansions}

In this section, we study the average weight of $F$-expansions of minimal 
weight.
For every $N\in\mathbb Z$, let $\|N\|_F$ be the weight of a corresponding 
$F$-expansion of minimal weight, i.e., $\|N\|_F=\|x\|$ if $x$ is an 
$F$-expansion of minimal weight with $x\sim_F N$.

\begin{theorem}
For positive integers $M$, we have, as $M\to\infty$,
$$
\frac1{2M+1}\sum_{N=-M}^M\|N\|_F=\frac15\,\frac{\log M}{\log\frac{1+\sqrt5}2}+
\mathcal O(1).
$$
\end{theorem}

\begin{proof}
Consider first $M=G_n$ for some $n>0$, where $G_n$ is defined as in the proof
of Proposition~\ref{Heuberger}, and let $W_n$ be the set of words 
$x=x_1\cdots x_n\in\{-1,0,1\}^n$ avoiding $11,101,1001,1\bar1,10\bar1$, and 
their opposites.
Then we have
$$
\frac1{2G_n+1}\sum_{N=-G_n}^{G_n}\|N\|_F=\frac1{\#W_n}\sum_{x\in W_n}\|x\|=
\sum_{j=1}^n\mathbf E\,X_j,
$$
where $\mathbf E\,X_j$ is the expected value of the random variable $X_j$ 
defined by
$$
\mathrm{Pr}[X_j=1]=\frac{\#\{x_1\cdots x_n\in W_n:x_j\ne 0\}}{\#W_n},
\mathrm{Pr}[X_j=0]=\frac{\#\{x_1\cdots x_n\in W_n:x_j=0\}}{\#W_n}
$$
Instead of $(X_j)_{1\le j\le n}$, we consider the sequence of random variables 
$(Y_j)_{1\le j\le n}$ defined by 
\begin{multline*}
\mathrm{Pr}[Y_1=y_1y_2y_3,\ldots,Y_j=y_jy_{j+1}y_{j+2}] \\
=\#\{x_1\cdots x_{n+2}\in W_n00:\,x_1\cdots x_{j+2}=y_1\cdots y_{j+2}\}/\#W_n,
\end{multline*}
$\mathrm{Pr}[Y_{j-1}=xyz,Y_j=x'y'z']=0$ if $x'\ne y$ or $y'\ne z$.
It is easy to see that $(Y_j)_{1\le j\le n}$ is a Markov chain, where
the non-trivial transition probabilities are given by
\begin{gather*}
1-\mathrm{Pr}[Y_{j+1}=000\mid Y_j=100]=\mathrm{Pr}[Y_{j+1}=00\bar1\mid Y_j=100]
=\frac{G_{n-j-2}-G_{n-j-3}}{G_{n-j+1}-G_{n-j}}, \\
1-2\,\mathrm{Pr}[Y_{j+1}=001\mid Y_j=000]=\mathrm{Pr}[Y_{j+1}=000\mid Y_j=000]=
\frac{2G_{n-j-3}+1}{2G_{n-j-2}+1},
\end{gather*}
and the opposite relations.
Since $G_n=c\beta^n+\mathcal O(1)$ (with $\beta=\frac{1+\sqrt5}2$, 
$c=\beta^3/5$), the transition probabilities satisfy 
$\mathrm{Pr}[Y_{j+1}=v\mid Y_j=u]=p_{u,v}+\mathcal O(\beta^{-n+j})$ with
$$
(p_{u,v})_{u,v\in\{100,010,001,000,00\bar1,0\bar10,00\bar1\}} = 
\begin{pmatrix}
0 & 0 & 0 & \frac2{\beta^2} & \frac1{\beta^3} & 0 & 0 \\
1 & 0 & 0 & 0 & 0 & 0 & 0 \\
0 & 1 & 0 & 0 & 0 & 0 & 0 \\
0 & 0 & \frac1{2\beta^2} & \frac1\beta & \frac1{2\beta^2} & 0 & 0 \\
0 & 0 & 0 & 0 & 0 & 1 & 0 \\
0 & 0 & 0 & 0 & 0 & 0 & 1 \\
0 & 0 & \frac1{\beta^3} & \frac2{\beta^2} & 0 & 0 & 0 
\end{pmatrix}.
$$ 
The eigenvalues of this matrix are $1,\frac{-1}\beta,\frac{\pm i}\beta, 
\frac{1\pm\,i\sqrt3}{2\beta},\frac{-1}{\beta^2}$. 
The stationary distribution vector (given by the left eigenvector to the 
eigenvalue~$1$) is $(\frac1{10},\frac1{10},\frac1{10},\frac25,
\frac1{10},\frac1{10},\frac1{10})$, thus we have
$$
\mathbf E\,X_j=\mathrm{Pr}[Y_j=100]+\mathrm{Pr}[Y_j=\bar100]=
1/5+\mathcal O\big(\beta^{-\min(j,n-j)}\big),
$$
cf.~\cite{DrmotaSteiner02}.
This proves the theorem for $M=G_n$.

If $G_n<M\le G_{n+1}$, then we have $\|N\|_F=1+\|N-F_n\|_F$ if $G_n<N\le M$, 
and a similar relation for $-M\le N<-G_n$.
With $G_n+1-F_n=-G_{n-2}$, we obtain 
\begin{multline*}
\sum_{N=-M}^M\|N\|_F = \sum_{N=-G_n}^{G_n}\|N\|_F+
\sum_{N=-G_{n-2}}^{M-F_n}(1+\|N\|_F)+\sum_{N=F_n-M}^{G_{n-2}}(1+\|N\|_F) \\
= \sum_{N=-G_n}^{G_n}\|N\|_F+\sum_{N=-G_{n-2}}^{G_{n-2}}\|N\|_F+
\mathrm{sgn}(M-F_n)\sum_{N=-|M-F_n|}^{|M-F_n|}\|N\|_F+\mathcal O(M) \\
= \frac2{5\log\beta}\big(F_n\log M+(M-F_n)\log|M-F_n|\big)+\mathcal O(M)= 
\frac{2M\log M}{5\log\beta}+\mathcal O(M)
\end{multline*}
by induction on $n$ and using $\frac{M-F_n}M\log|\frac{M-F_n}M|=\mathcal O(1)$.
\end{proof}

\begin{remark}
As in~\cite{DrmotaSteiner02}, a central limit theorem for the distribution of $\|N\|_F$ can be proved, even if we restrict the numbers $N$ to polynomial sequences or prime numbers.
\end{remark}

\begin{remark}
If we partition the interval 
$\big[\frac{-\beta^2}{\beta^2+1},\frac{\beta^2}{\beta^2+1}\big)$, where the
transformation 
$\tau:z\mapsto\beta z-\big\lfloor\frac{\beta^2+1}{2\beta}z+1/2\big\rfloor$
of the proof of Proposition~\ref{particular} is defined,
into intervals 
$I_{\bar100}=\big[\frac{-\beta^2}{\beta^2+1},\frac{-\beta}{\beta^2+1}\big)$, 
$I_{0\bar10}=\big[\frac{-\beta}{\beta^2+1},\frac{-1}{\beta^2+1}\big)$, 
$I_{00\bar1}=\big[\frac{-1}{\beta^2+1},\frac{-1/\beta}{\beta^2+1}\big)$, 
$I_{000}=\big[\frac{-1/\beta}{\beta^2+1},\frac{1/\beta}{\beta^2+1}\big)$, 
$I_{001}=\big[\frac{1/\beta}{\beta^2+1},\frac1{\beta^2+1}\big)$, 
$I_{010}=\big[\frac1{\beta^2+1},\frac\beta{\beta^2+1}\big)$, 
$I_{100}=\big[\frac\beta{\beta^2+1},\frac{\beta^2}{\beta^2+1}\big)$, then we 
have $p_{u,v}=\lambda(\tau(I_u)\cap I_v)/\lambda(\tau(I_u))$, where $\lambda$
denotes the Lebesgue measure.
\end{remark}

\section{Tribonacci case}

In this section, let $\beta>1$ be the Tribonacci number,
$\beta^3=\beta^2+\beta+1$ ($\beta\approx 1.839$).
Since $1=\decdot 111$, we have $2=10\decdot001$ and $\beta$ satisfies ($\mathrm D_2$).
Here, the digits of arbitrary $\beta$-expansions of minimal weight are in $\{-5,\ldots,5\}$ since $6=1000\decdot00\bar10\bar10\bar1$.
We have $5=101\decdot100011$ and we will show that $101100011$ is a $\beta$-expansion of minimal weight, thus $5$ is also a $\beta$-expansion of minimal weight.

The proofs of the results in this section run along the same lines as in the 
Golden Ratio case.
Therefore we give only an outline of them.

\subsection{$\beta$-expansions of minimal weight}

\begin{figure}[ht]
\centerline{\includegraphics[scale=.94]{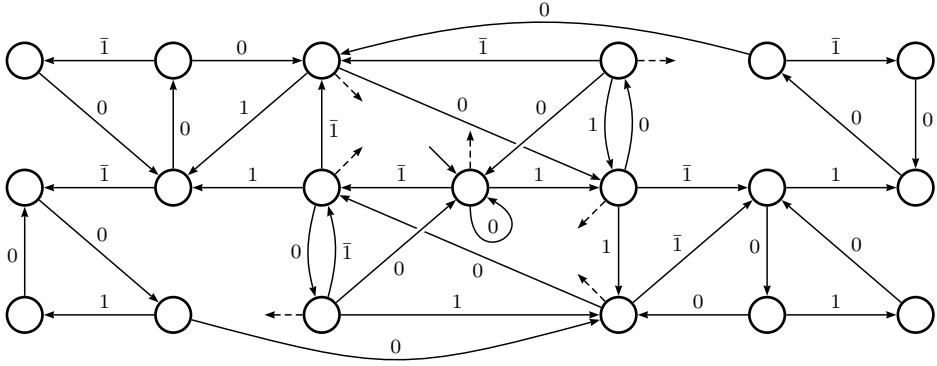}}
\caption{Automata $\mathcal M_\beta$, $\beta^3=\beta^2+\beta+1$, and 
$\mathcal M_T$.} \label{figexp2}
\end{figure}

\begin{figure}[ht]
\centerline{\includegraphics{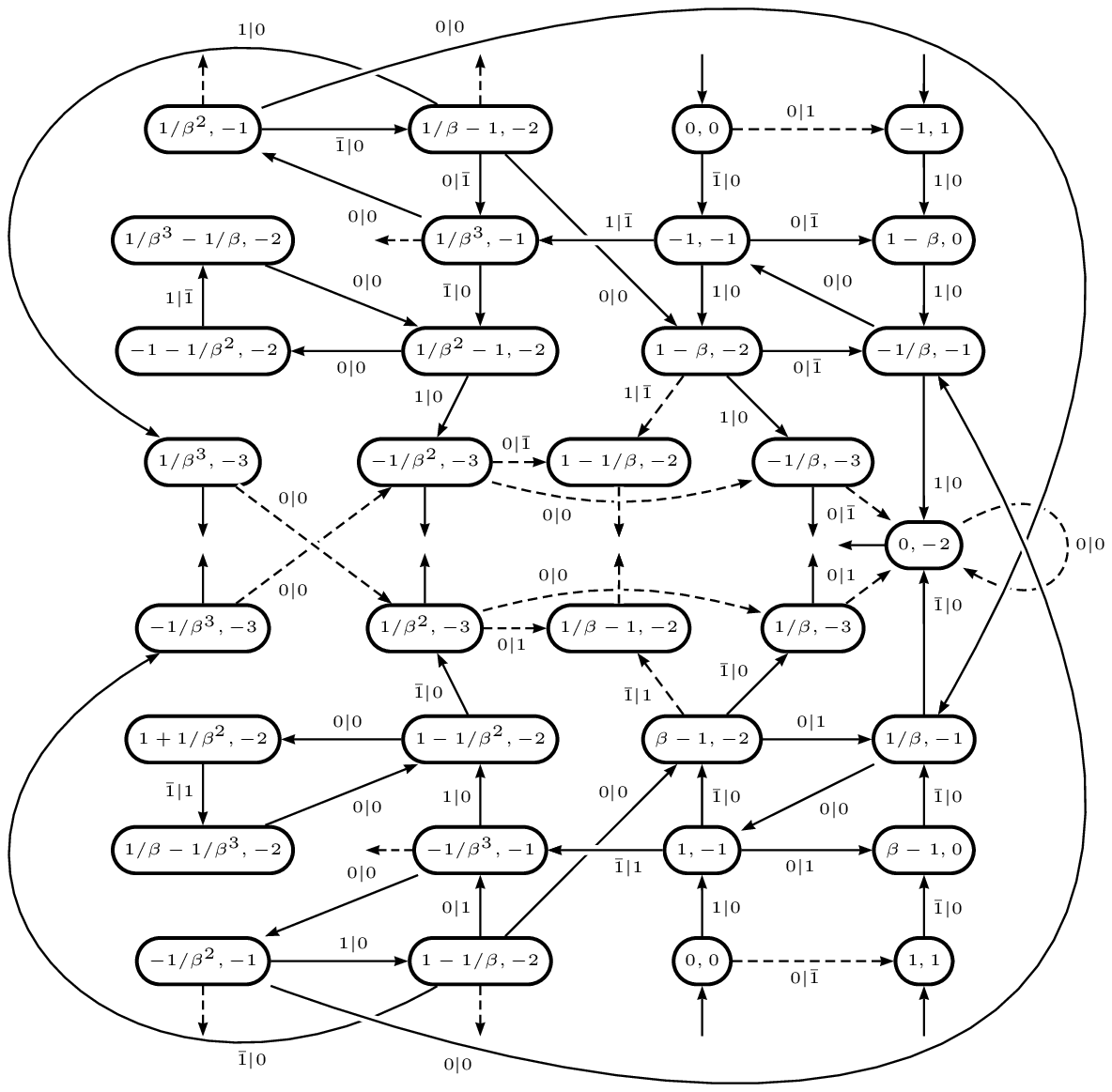}}
\caption{The relevant part of $\mathcal S_\beta$, $\beta^3=\beta^2+\beta+1$, 
and $\mathcal S_T$.} \label{figforb2}
\end{figure}

All words which are not accepted by the automaton $\mathcal M_\beta$ in
Figure~\ref{figexp2}, where all states are terminal, are $\beta$-heavy since
they contain a factor which is accepted by the input automaton of the transducer in Figure~\ref{figforb2} (without the dashed arrows).

\begin{proposition}\label{particulartribo}
If $\beta>1$ is the Tribonacci number, then every $z\in\mathbb R$ has a 
$\beta$-expansion of the form $z=y_1\cdots y_k\decdot y_{k+1}y_{k+2}\cdots$ 
with $y_j\in\{-1,0,1\}$ such that $y_1y_2\cdots$ avoids the set 
$X=\{ 11, 101, 1\bar1,\mbox{ and their opposites}\}$.
If $z\in\mathbb Z[\beta]=\mathbb Z[\beta^{-1}]$, then this expansion is unique 
up to leading zeros.
\end{proposition}

The expansion in Proposition~\ref{particulartribo} is given by the 
transformation
$$
\tau:\ \left[\frac{-\beta}{\beta+1},\frac\beta{\beta+1}\right)\to
\left[\frac{-\beta}{\beta+1},\frac\beta{\beta+1}\right),\quad 
\tau(z)=\beta z-\left\lfloor\frac{\beta+1}2z+\frac12\right\rfloor.
$$
Note that the word avoiding $X$ with maximal value is $(100)^\omega$, 
$\decdot(100)^\omega=\frac\beta{\beta+1}$.

\begin{remark}
The transformation
$\tau(z)=\beta z-\big\lfloor\frac{\beta^2-1}2z+\frac12\big\rfloor$ on
$\big[\frac{-\beta}{\beta^2-1},\frac\beta{\beta^2-1}\big)$ provides a unique 
expansion avoiding the factors $11$, $1\bar1$, $10\bar1$ and their opposites. 
\end{remark}

\begin{figure}[ht]
\centerline{\includegraphics[scale=.9]{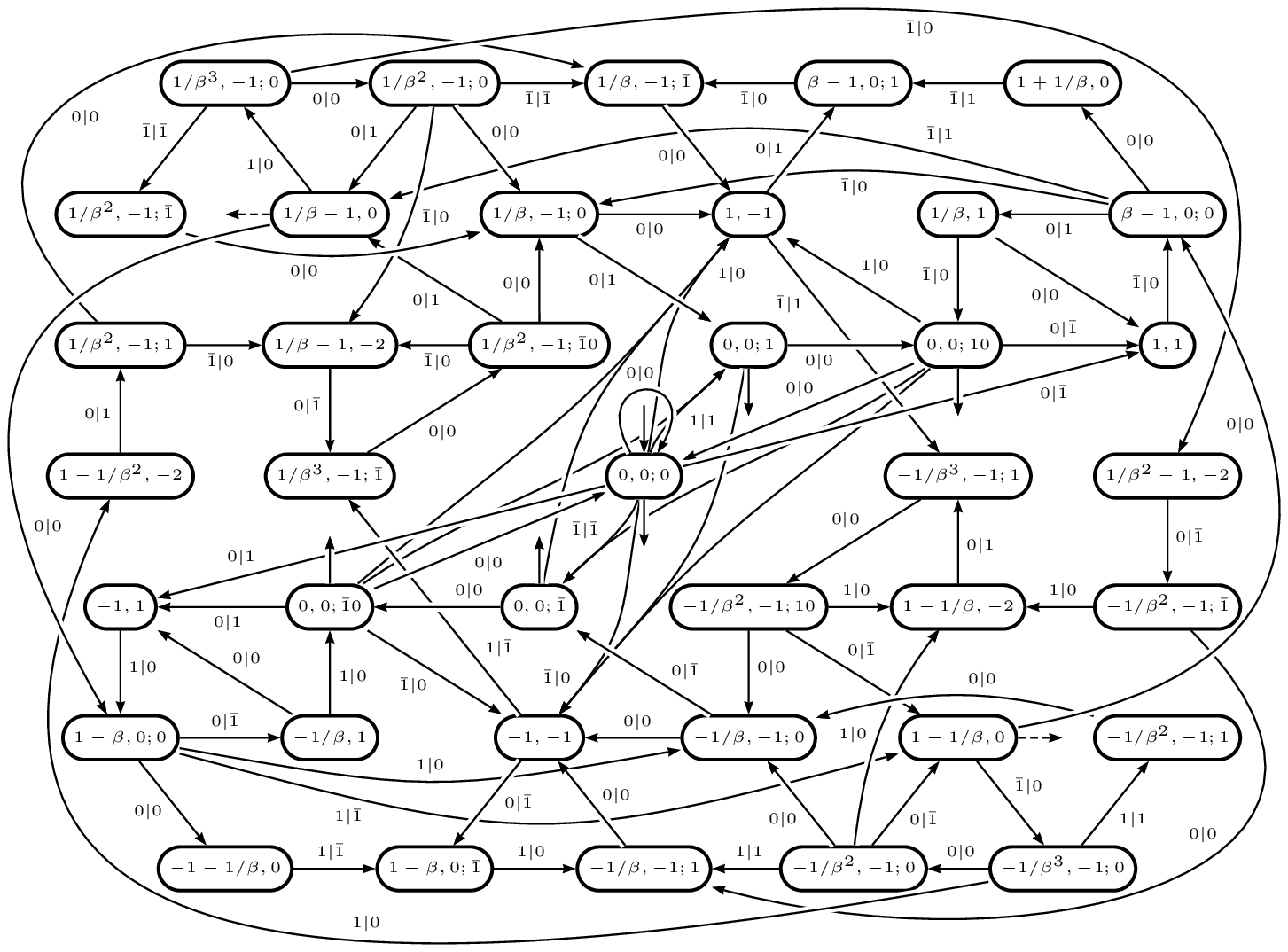}}
\caption{Normalizing transducer $\mathcal N_\beta$, $\beta^3=\beta^2+\beta+1$.} 
\label{normaltribo}
\end{figure}

\begin{proposition}
The conversion of an arbitrary expansion accepted by the automaton $\mathcal M_\beta$ in Figure~\ref{figexp2} into the expansion avoiding $X=\{ 11, 101, 1\bar1,\mbox{ and their opposites}\}$ is realized by the transducer $\mathcal N_\beta$ in Figure~\ref{normaltribo} and does not change the weight.
\end{proposition}

\begin{theorem}
If $\beta$ is the Tribonacci number, then the set of $\beta$-expansions of 
minimal weight in $\{-1,0,1\}^*$ is recognized by the finite automaton 
$\mathcal M_\beta$ of Figure~\ref{figexp2} where all states are terminal.
\end{theorem}

\subsection{Branching transformation}

\begin{figure}[ht]
\centerline{\includegraphics[scale=.9]{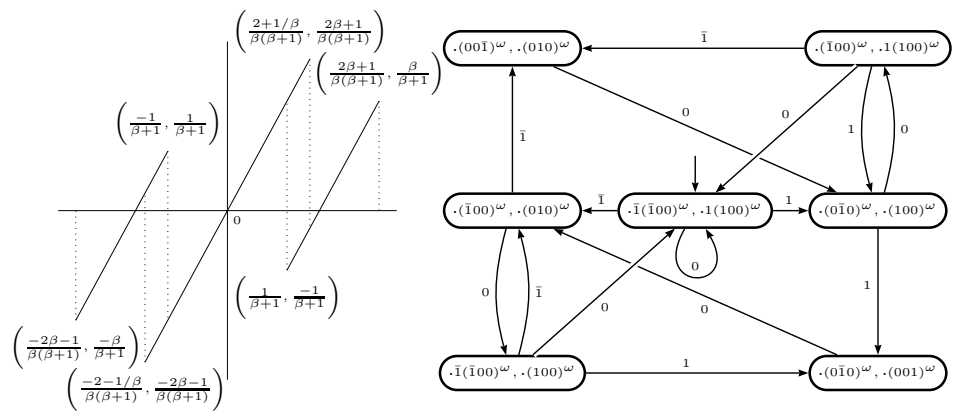}}
\caption{Branching transformation, corresponding automaton, 
$\beta^3=\beta^2+\beta+1$.} \label{figbranchtribo}
\end{figure}

Contrary to the Golden Ratio case, we cannot obtain all $\beta$-expansions of 
minimal weight by the help of a piecewise linear branching transformation:
If $z=\decdot01(001)^n$, then we have no $\beta$-expansion of minimal weight
of the form $z=\decdot1x_2x_3\cdots$, whereas $z'=\decdot0011$ has the 
expansion $\decdot1\bar1$, and $z'<z$.
On the other hand, $z=\decdot1(100)^n11$ has no $\beta$-expansion of minimal 
weight of the form $z=\decdot1x_2x_3\cdots$ (since $1(100)^n11$ is 
$\beta$-heavy but $(100)^n11$ is not $\beta$-heavy), whereas $z'=\decdot1101$ 
is a $\beta$-expansion of minimal weight, and $z'>z$.
Hence the maximal interval for the digit $1$ is 
$[\decdot(010)^\omega,\decdot1(100)^\omega]$, with 
$\decdot(010)^\omega=\frac\beta{\beta^3-1}=\frac1{\beta+1}$ and
$\decdot1(100)^\omega=\frac{2\beta+1}{\beta(\beta+1)}$.
The corresponding branching transformation and the possible expansions
are given in Figure~\ref{figbranchtribo}.

\subsection{Tribonacci numeration system}

The linear numeration system canonically associated with the Tribonacci number 
is the Tribonacci numeration system defined by the sequence $T=(T_n)_{n\ge 0}$ 
with $T_0=1$, $T_1=2$, $T_2=4$, and $T_n=T_{n-1}+T_{n-2}+T_{n-3}$ for $n\ge3$.
Any non-negative integer $N<T_n$ has a representation 
$N=\sum_{j=1}^n x_jT_{n-j}$ with the property that $x_1\cdots x_n\in\{0,1\}^*$ 
does not contain the factor $111$.
The relation $\sim_T$ and the properties \emph{$T$-heavy}, 
\emph{$T$-expansion of minimal weight} and \emph{strongly $T$-heavy} are 
defined analogously to the Fibonacci numeration system.
We have $20^n\sim_T 100010^{n-3}$ for $n\ge3$, $200\sim_T 1001$, $20\sim_T 100$
and $2\sim_T 10$, therefore for every $x\in\mathbb Z^*$ there exists some
$y\in\{-1,0,1\}^*$ with $x\sim_T y$ and $\|y\|\le\|x\|$. 
Since the difference of $1(0\bar10)^{n/3}$ and $(100)^{n/3}$ is 
$1(\bar1\bar10)^{n/3}\sim_T 1$, we obtain the following proposition.

\begin{proposition}\label{particularinttribo}
Every $N\in\mathbb Z$ has a unique representation $N=\sum_{j=1}^n y_jT_{n-j}$
with $y_1\ne0$ and $y_1\cdots y_n\in\{-1,0,1\}^*$ avoiding 
$X=\{ 11, 101, 1\bar1,\mbox{ and their opposites}\}$.
\end{proposition}

If $z=a_1\cdots a_n\decdot=m_2m_1m_0\decdot$, then 
$N=\sum_{j=1}^n a_jT_{n-j}=4m_2+2m_1+m_0=0$ if and only if $m_0=2m_0'$ and 
$m_1=-2m_2-m_0'$, i.e., $z=-m_2/\beta^2+m_0'/\beta^3$, hence all states 
$s=m/\beta^2+m'/\beta^3$ with some $m,m'\in\mathbb Z$ are terminal states in
the redundancy transducer $\mathcal R_T$.
The transducer $\mathcal S_T$, which is given by Figure~\ref{figforb2} 
including the dashed arrows except that the states $(\pm1/\beta,-3)$ are not 
terminal, shows that all strictly $\beta$-heavy words in $\{-1,0,1\}^*$ are 
strongly $T$-heavy, but that some other $x\in\{-1,0,1\}^*$ are $T$-heavy as 
well.
Thus the $T$-expansions of minimal weight are a subset of the set recognized by 
the automaton $\mathcal M_\beta$ in Figure~\ref{figexp2}.
Every set $Q_u$ and $Q_u'$, $u\in\{0,1,10,11\}$, contains a terminal state
$(0,0;w)$ or $(1-1/\beta,0)$, hence the words labelling paths ending in these
states are $T$-expansions of minimal weight.
The sets $Q_u$ and $Q_u'$, $u\in\{1\bar1,1\bar10,1\bar11,1\bar10,1\bar101\}$,
contain states $(\pm1/\beta^3,-1;w)$, $(\pm1/\beta^2,-1;w)$, 
$(\pm(1-1/\beta),-2)$, hence the words labelling paths ending in these states 
are $T$-heavy, and we obtain the following theorem.

\begin{theorem}
The $T$-expansions of minimal weight in $\{-1,0,1\}^*$ are exactly the words 
which are accepted by $\mathcal M_T$, which is the automaton in 
Figure~\ref{figexp2} where only the states with a dashed outgoing arrow are 
terminal. 
The words given by Proposition~\ref{particularinttribo} are $T$-expansions of
minimal weight.
\end{theorem}

\subsection{Weight of the expansions}

Let $W_n$ be the set of words $x=x_1\cdots x_n\in\{-1,0,1\}^n$ avoiding the 
factors $11$, $101$, $1\bar1$, and their opposites.
Then the sequence of random variables $(Y_j)_{1\le j\le n}$ defined by
$$
\mathrm{Pr}[Y_1=y_1y_2,\ldots,Y_j=y_jy_{j+1}]=\frac{\#\{x_1\cdots x_{n+1}\in 
W_n0:\,x_1\cdots x_{j+1}=y_1\cdots y_{j+1}\}}{\#W_n}
$$
is Markov with transition probabilities 
$\mathrm{Pr}[Y_{j+1}=v\mid Y_j=u]=p_{u,v}+\mathcal O(\beta^{-n+j})$,
$$
(p_{u,v})_{u,v\in\{10,01,00,0\bar1,\bar10\}} = 
\begin{pmatrix}
0 & 0 & \frac{\beta^2-1}{\beta^2} & \frac1{\beta^2} & 0 \\
1 & 0 & 0 & 0 & 0 \\
0 & \frac{\beta-1}{2\beta} & \frac1\beta & \frac{\beta-1}{2\beta} & 0\\
0 & 0 & 0 & 0 & 1 \\
0 & \frac1{\beta^2} & \frac{\beta^2-1}{\beta^2} & 0 & 0 
\end{pmatrix}.
$$ 
The eigenvalues of this matrix are 
$1,\pm\frac1\beta,\frac{-\beta-1\pm\,i\sqrt{3\beta^3-\beta}}{2\beta^3}$,
and the stationary distribution vector of the Markov chain is 
$\big(\frac{\beta^3/2}{\beta^5+1},\frac{\beta^3/2}{\beta^5+1}, 
\frac{\beta^3+\beta^2}{\beta^5+1},\frac{\beta^3/2}{\beta^5+1},
\frac{\beta^3/2}{\beta^5+1}\big)$.
We obtain the following theorem 
(with $\frac{\beta^3}{\beta^5+1}=\decdot(0011010100)^\omega\approx0.28219$).

\begin{theorem}
For positive integers $M$, we have, as $M\to\infty$,
$$
\frac1{2M+1}\sum_{N=-M}^M\|N\|_T=
\frac{\beta^3}{\beta^5+1}\,\frac{\log M}{\log\beta}+\mathcal O(1).
$$
\end{theorem}

\section{Smallest Pisot number case}\label{sp}

The smallest Pisot number $\beta \approx 1.325$ satisfies $\beta^3=\beta+1$.
Since $1=\decdot011=\decdot10001$ implies $2=100\decdot00001$ as well as 
$2=1000\decdot000\bar1$, ($\mathrm D_2$) holds.
We have furthermore $3=\beta^4-\beta^{-9}$, thus all $\beta$-expansions of minimal weight have digits in $\{-2,\ldots,2\}$.

\subsection{$\beta$-expansions of minimal weight}

\begin{figure}[ht]
\centerline{\includegraphics[scale=.9]{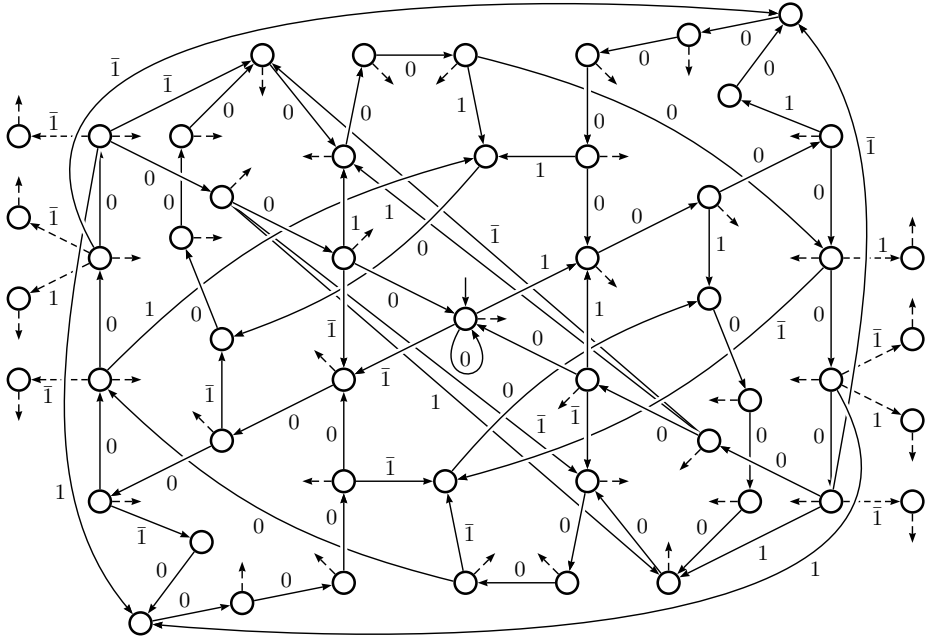}}
\caption{Automata $\mathcal M_\beta$, $\beta^3=\beta+1$, and $\mathcal M_S$.} 
\label{figexpminpisot}
\end{figure}

\begin{figure}[ht]
\centerline{\includegraphics{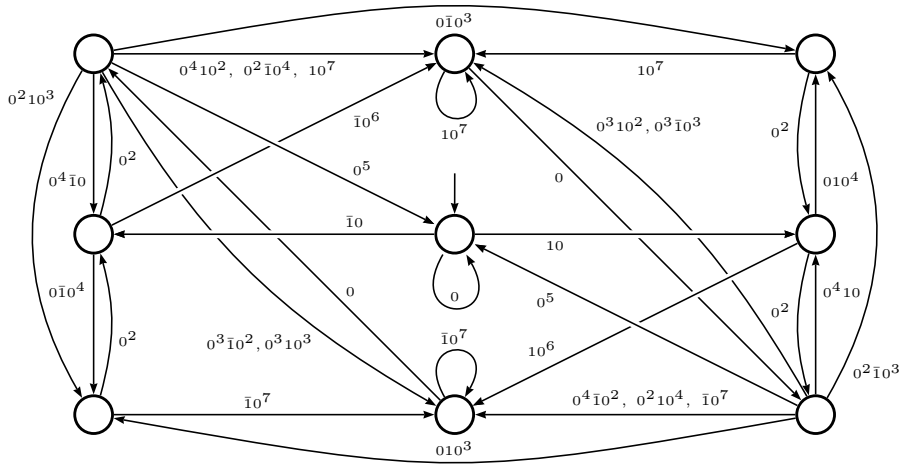}}
\caption{Compact representation of $\mathcal M_\beta$.}
\end{figure}

\begin{figure}[ht]
\centerline{\includegraphics[scale=.9]{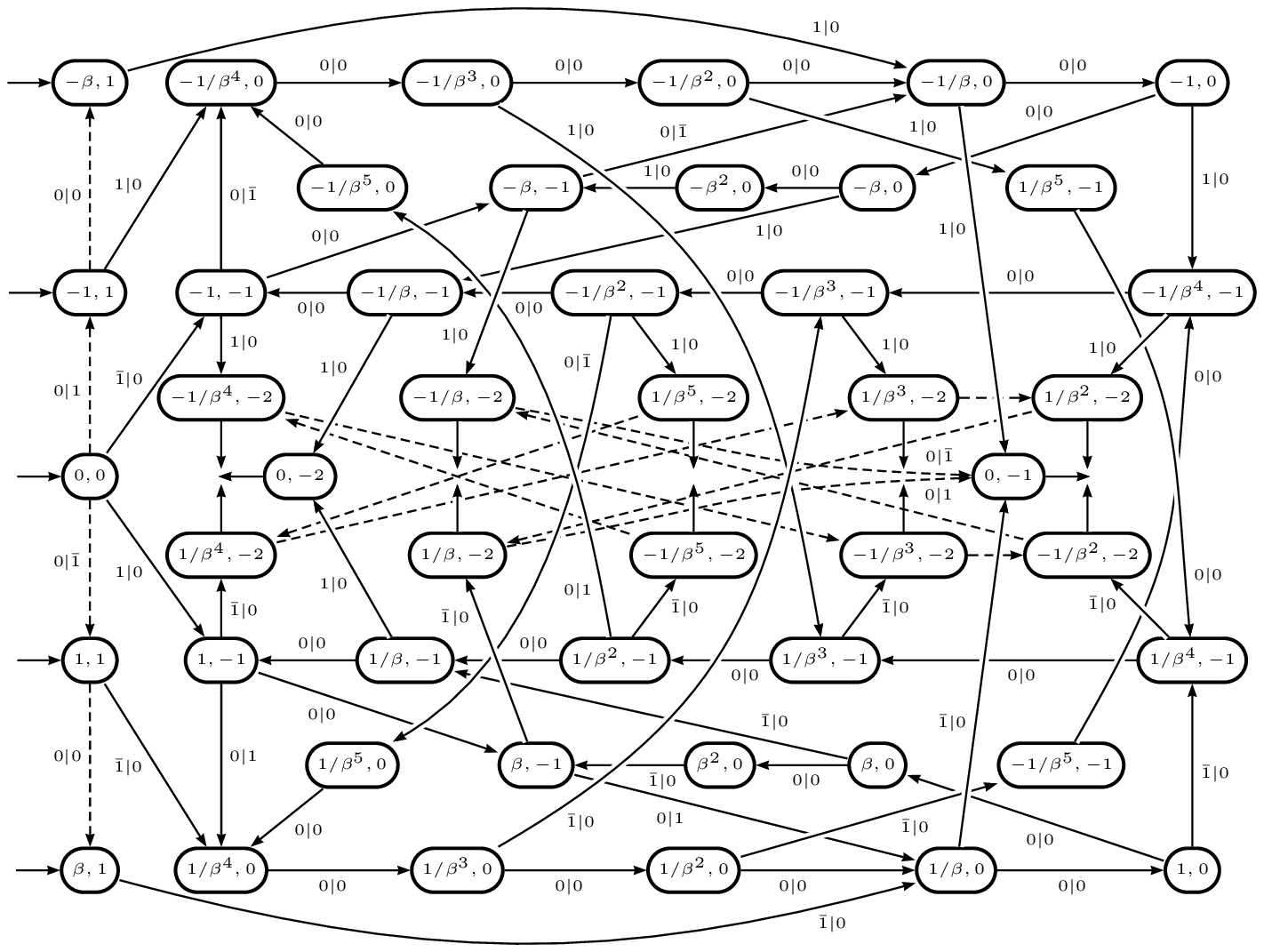}}
\caption{The relevant part of $\mathcal S_\beta$, $\beta^3=\beta+1$.} 
\label{figforbminpisot}
\end{figure}

Let $\mathcal M_\beta$ be the automaton in Figure~\ref{figexpminpisot} without 
the dashed arrows where all states are terminal.
Then it is a bit more difficult than in the Golden Ration and the Tribonacci cases to see that all words which are not accepted by $\mathcal M_\beta$ are $\beta$-heavy, not only because the automata are larger but also because some inputs of the transducer in Figure~\ref{figforbminpisot} are not strictly $\beta$-heavy (but of course still $\beta$-heavy). 
We refer to \cite{FrougnySteiner08} for details.

\begin{proposition}\label{particularminpisot}
If $\beta$ is the smallest Pisot number, then every $z\in\mathbb R$ has a 
$\beta$-expansion of the form $z=y_1\cdots y_k\decdot y_{k+1}y_{k+2}\cdots$ 
with $y_j\in\{-1,0,1\}$ such that $y_1y_2\cdots$ avoids the set 
$X=\{10^61, 10^k1, 10^k\bar1, 0\le k\le 5,\mbox{ and their opposites}\}$.
If $z\in\mathbb Z[\beta]=\mathbb Z[\beta^{-1}]$, then this expansion is unique 
up to leading zeros.
\end{proposition}

The expansion in Proposition~\ref{particularminpisot} is given by the transformation
$$
\tau:\ \Big[\frac{-\beta^3}{\beta^2+1},\frac{\beta^3}{\beta^2+1}\Big)\to
\Big[\frac{-\beta^3}{\beta^2+1},\frac{\beta^3}{\beta^2+1}\Big),\quad 
\tau(x)=\beta x-\Big\lfloor\frac{\beta^2+1}{2\beta^2}x+\frac12\Big\rfloor
$$
since 
$\tau\big[\frac{\beta^2}{\beta^2+1},\frac{\beta^3}{\beta^2+1}\big)=
\big[\frac{\beta^3}{\beta^2+1}-1,\frac{\beta^4}{\beta^2+1}-1\big)=
\big[-\frac{1/\beta^3}{\beta^2+1},\frac{1/\beta^4}{\beta^2+1}\big)$.
The word avoiding $X$ with maximal value is $(10^7)^\omega$,
$\decdot(10^7)^\omega=\beta^7/(\beta^8-1)=\beta^3/(\beta^2+1)$.

\begin{remark}
The transformation 
$\tau(z)=\beta z-\big\lfloor\frac1\beta z+\frac12\big\rfloor$ on
$\big[-\frac{\beta^2}2,\frac{\beta^2}2\big)$ provides a unique 
expansion avoiding $10^6\bar1$ instead of $10^61$.
\end{remark}

\begin{figure}[ht]
\centerline{\includegraphics[scale=.88]{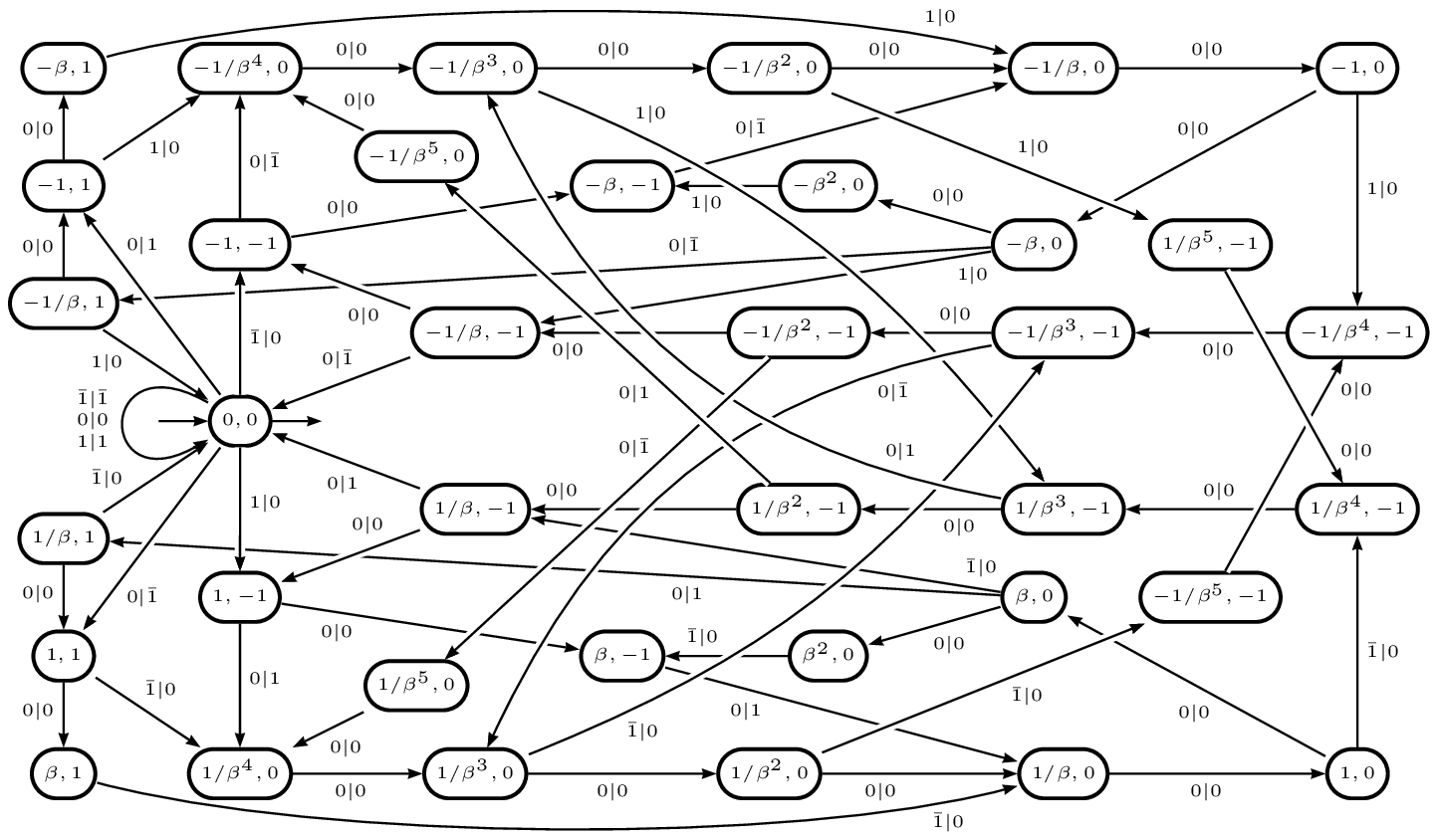}}
\caption{Transducer $\mathcal N_\beta$ normalizing $\beta$-expansions of 
minimal weight, $\beta^3=\beta+1$.} \label{normalminpisot}
\end{figure}

\begin{proposition}
The conversion of an arbitrary expansion accepted by $\mathcal M_\beta$ into the expansion avoiding $X=\{10^61, 10^k1, 10^k\bar1, 0\le k\le 5,\mbox{ and their opposites}\}$ is realized by the transducer $\mathcal N_\beta$ in Figure~\ref{normalminpisot} and does not change the weight.
\end{proposition}

\begin{theorem}
If $\beta$ is the smallest Pisot number, then the set of $\beta$-expansions of 
minimal weight in $\{-1,0,1\}^*$ is recognized by the finite automaton 
$\mathcal M_\beta$ of Figure~\ref{figexpminpisot} (without the dashed arrows)
where all states are terminal.
\end{theorem}

\subsection{Branching transformation}

\begin{figure}[ht]
\centerline{\includegraphics[scale=.9]{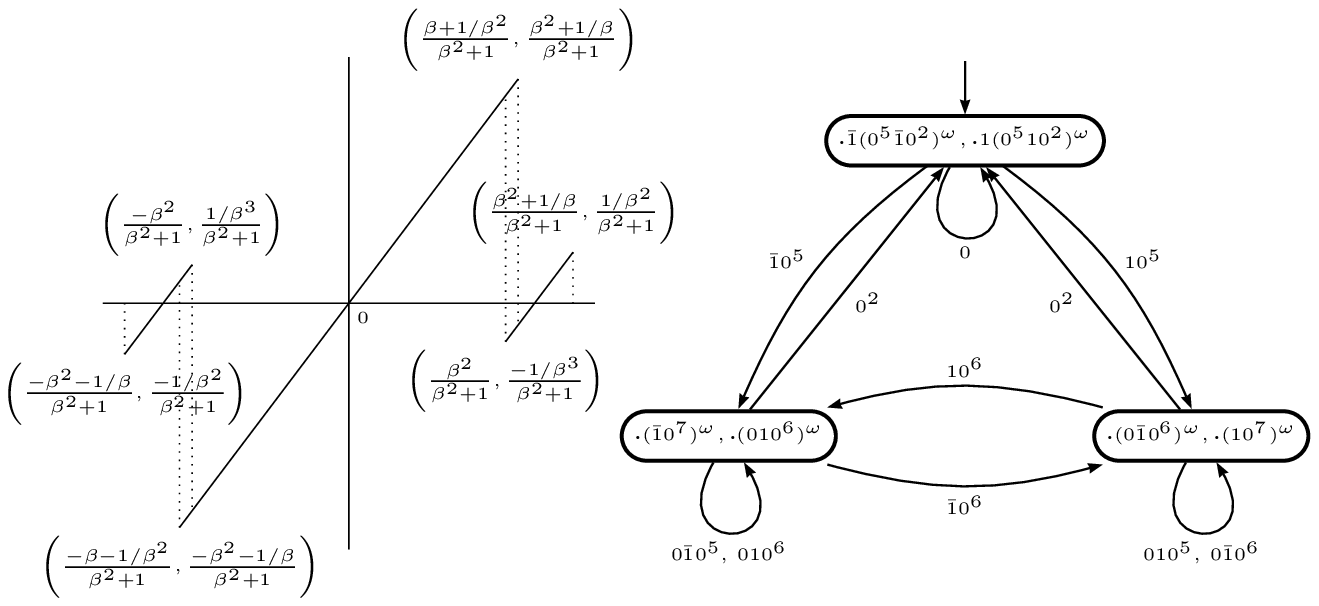}}
\caption{Branching transformation and corresponding automaton, 
$\beta^3=\beta+1$.} \label{figbranchminpisot}
\end{figure}

In the case of the smallest Pisot number $\beta$, the 
maximal interval for the digit $1$ is 
$[\decdot(010^6)^\omega,\decdot1(0^510^2)^\omega]$, with
$\decdot(010^6)^\omega=\frac{\beta^2}{\beta^2+1}$ and
$\decdot1(0^510^2)^\omega=\frac{\beta^2+1/\beta}{\beta^2+1}$.
The corresponding branching transformation and expansions
are given in Figure~\ref{figbranchminpisot}.

\subsection{Integer expansions}

Let $(S_n)_{n\ge0}$ be a linear numeration system associated with the 
smallest Pisot number $\beta$ which is defined as follows:
$$
S_0=1,\;S_1=2,\;S_2=3,\;S_3=4,\quad S_n=S_{n-2}+S_{n-3}\;\mbox{ for }n\ge 4.
$$
Note that we do not choose the canonical numeration system associated with the 
smallest Pisot number, which is defined by 
$U_0=1,U_1=2,U_2=3,U_3=4,U_4=5,U_n=U_{n-1}+U_{n-5}$ for $n\ge5$, since 
$U_n=U_{n-2}+U_{n-3}$ holds only for $n\equiv 1\bmod 3$, $n\ge4$.

For every $x\in\mathbb Z^*$, there exists $y\in\{-1,0,1\}^*$ with $x\sim_S y$, 
$\|y\|\le \|x\|$, since $2\sim_S 10$, $20\sim_S 1000$, $200\sim_S 1010$, 
$20^3\sim_S 10100$, $20^4\sim_S 100100$, $20^5\sim_S 1010^4$, 
$20^n\sim_S 10^610^{n-5}$ for $n\ge6$.

\begin{proposition}\label{particularintminpisot}
Every $N\in\mathbb Z$ has a unique representation $N=\sum_{j=1}^n y_jS_{n-j}$
with $y_1\ne0$ and $y_1\cdots y_n\in\{-1,0,1\}^*$ avoiding the set 
$X=\{10^61, 10^k1, 10^k\bar1, 0\le k\le 5$, and their opposites$\}$, with the
exception that $10^61,10^51,10^5\bar1,10^4\bar1$ and their opposites are 
possible suffixes of $y_1\cdots y_n$.
\end{proposition}

As for the Fibonacci numeration system, Proposition~\ref{particularintminpisot} is proved by considering $g_n$, the smallest positive integer with an expansion of length $n$ starting with $1$ avoiding these factors, and $G_n$, the largest integer of this kind.
The representations of $g_{n+1}$ and $G_n$, $n\ge1$, depending on the 
congruence class of $n$ modulo $8$ are given by the following table.

$$
\begin{array}{c|c|c|c}
n\equiv j\bmod 8 & g_{n+1} & G_n & g_{n+1}-G_n \\ \hline
1,2,3,4 & 1(0^6\bar10)^{n/8} & (10^7)^{n/8} & 1\bar10^{j-1}\sim_S 1 \\
5 & 1(0^6\bar10)^{(n-5)/8}0^4\bar1 & (10^7)^{(n-5)/8}10^4 & 1\bar1000\bar1\sim_S 1 \\
6 & 1(0^6\bar10)^{(n-6)/8}0^5\bar1 & (10^7)^{(n-6)/8}10^5 & 1\bar10000\bar1\sim_S 1\bar1\sim_S 1 \\
7 & 1(0^6\bar10)^{(n-7)/8}0^6\bar1 & (10^7)^{(n-7)/8}10^51 & 1\bar100000\bar2\sim_S 10\bar2\sim_S 1 \\
0 & 1(0^6\bar10)^{n/8} & (10^7)^{n/8-1}10^61 & 1\bar100000\bar1\bar1\sim_S 10\bar1\bar1\sim_S 1
\end{array}
$$
For the calculation of $g_{n+1}-G_n$ we have used 
$S_n-S_{n-1}-S_{n-7}=S_{n-8}$ for $n\ge9$.

Since $S_n=S_{n-2}-S_{n-3}$ holds only for $n\ge4$ and not for $n=3$, 
determining when $x\sim_S y$ is more complicated than for $\sim_F$ and 
$\sim_T$. 
If $z=a_1\cdots a_n\decdot=m_3m_2m_1a_n\decdot$, then we have 
$N=\sum_{j=1}^n a_jS_{n-j}=4m_3+3m_2+2m_1+a_n$.
We have to distinguish between different values of $a_n$.
\begin{itemize}
\item
If $a_n=0$, then $N=0$ if and only if $m_2=2m_2'$, $m_1=-2m_3-3m_2'$, hence 
$$
z=m_3(\beta^3-2\beta)+m_2'(2\beta^2-3\beta)=
-m_3/\beta^4-m_2'(1/\beta^4+1/\beta^7).
$$
In particular, $m_2'=0,m_3\in\{0,\pm1\}$ implies $N=0$ if 
$z\in\{0,\pm1/\beta^4\}$.
\item
If $a_n=1$, then $N=0$ if and only if $m_2=2m_2'-1$, $m_1=-2m_3-3m_2'+1$, 
$$
z=m_3(\beta^3-2\beta)+m_2'(2\beta^2-3\beta)-\beta^2+\beta+1=
-m_3/\beta^4-m_2'(1/\beta^4+1/\beta^7)+1/\beta^2.
$$
In particular, $m_3m_2'\in\{00,\bar11,01\}$ provides $N=0$ if 
$z\in\{1/\beta^2,1/\beta^3,1/\beta^5\}$.
\item
If $a_n=2$, then $m_3m_2m_1\in\{00\bar1,\bar101\}$ provides
$N=0$ if $z\in\{2-\beta,1\}$.
\end{itemize}
We have $x_1\cdots x_n\sim_S y_1\cdots y_n$ if the corresponding path in 
$\mathcal R_\beta$ ends in a state $z$ corresponding to $a_n=x_n-y_n$ (or in 
$-z$, $a_n=y_n-x_n$) and obtain the following theorem.

\begin{theorem}\label{integerminpisot}
The set of $S$-expansions of minimal weight in $\{-1,0,1\}^*$ is recognized by 
$\mathcal M_S$, which is the automaton in Figure~\ref{figexpminpisot} including 
the dashed arrows.
The words given by Proposition~\ref{particularintminpisot} are $S$-expansions 
of minimal weight.
\end{theorem}

For details on the proof of Theorem~\ref{integerminpisot}, we refer again to~\cite{FrougnySteiner08}.

\subsection{Weight of the expansions}

Let $W_n$ be the set of words $x=x_1\cdots x_n\in\{-1,0,1\}^n$ avoiding the 
factors given by Proposition~\ref{particularintminpisot}.
Then the sequence of random variables $(Y_j)_{1\le j\le n}$ defined by
\begin{multline*}
\mathrm{Pr}[Y_1=y_1\cdots y_7,\ldots,Y_j=y_j\cdots y_{j+6}] \\
=\#\{x_1\cdots x_{n+6}\in W_n0^6:\,x_1\cdots x_{j+6}=y_1\cdots y_{j+6}\}/\#W_n
\end{multline*}
is Markov with transition probabilities 
$\mathrm{Pr}[Y_{j+1}=v\mid Y_j=u]=p_{u,v}+\mathcal O(\beta^{-n+j})$,
$$
(p_{u,v})_{u,v\in\{10^6,\ldots,0^61,0^7,0^6\bar1,\ldots,\bar10^6\}} = 
\begin{pmatrix}
0 & \cdots & \cdots & 0 & \frac2{\beta^3} & \frac1{\beta^7} & 0 & \cdots & 0 \\
1 & \ddots & & \vdots & 0 & 0 & \vdots & & \vdots \\
0 & \ddots & \ddots & \vdots & \vdots & \vdots & \vdots & & \vdots \\
\vdots & \ddots & 1 & 0 & 0 & 0 & \vdots & & \vdots \\
\vdots & & 0 & \frac1{2\beta^5} & \frac1\beta & \frac1{2\beta^5} & 0 & & \vdots \\
\vdots & & \vdots & 0 & 0 & 0 & 1 & \ddots & \vdots \\
\vdots & & \vdots & \vdots & \vdots & \vdots & \ddots & \ddots & 0 \\
\vdots & & \vdots & 0 & 0 & \vdots & & \ddots & 1 \\
0 & \cdots & 0 & \frac1{\beta^7} & \frac2{\beta^3} & 0 & \cdots & \cdots & 0 
\end{pmatrix}.
$$
The left eigenvector to the eigenvalue $1$ of this matrix is 
$\frac1{14+4\beta^2}(1,\ldots,1,4\beta^2,1,\ldots,1)$, and we obtain the
following theorem (with $\frac1{7+2\beta^2}\approx0.09515$).

\begin{theorem}
For positive integers $M$, we have, as $M\to\infty$,
$$
\frac1{2M+1}\sum_{N=-M}^M\|N\|_S=
\frac1{7+2\beta^2}\,\frac{\log M}{\log\beta}+\mathcal O(1).
$$
\end{theorem}

\section{Concluding remarks}
Another example of a number $\beta<2$ of small degree satisfying ($\mathrm D_2$), which is not studied in this article, is the Pisot number satisfying $\beta^3=\beta^2+1$, with $2=100\decdot0000\bar1$.

\smallskip
A question which is not approached in this paper concerns $\beta$-expansions 
of minimal weight in $\{1-B,\ldots,B-1\}^*$ when $\beta$ does not satisfy ($\mathrm D_B$), in particular minimal weight expansions on the alphabet $\{-1,0,1\}$ when $\beta<3$ and ($\mathrm D_2$) does not hold.

\smallskip
In view of applications to cryptography, we present a summary of the average 
minimal weight of representations of integers in linear numeration systems 
$(U_n)_{n\ge 0}$ associated with different $\beta$, with digits in $A=\{0,1\}$ 
or in $A=\{-1,0,1\}$. 

$$
\begin{array}{c|c|c|c}
U_n & A & \beta & \text{average }\|N\|_U\text{ for }N\in\{-M,\ldots,M\}\\
\hline\hline
2^n & \{0,1\} & 2 & (\log_2M)/2 \\ \hline
2^n & \{-1,0,1\} & 2 & (\log_2M)/3 \\ \hline
F_n & \{0,1\} & \frac{1+\sqrt5}2 & 
(\log_\beta M)/(\beta^2+1)\approx0.398\log_2M \\ \hline
F_n & \{-1,0,1\} & \frac{1+\sqrt5}2 & 
(\log_\beta M)/5\approx0.288\log_2M \\ \hline
T_n & \{-1,0,1\} & \beta^3=\beta^2+\beta+1 & 
(\log_\beta M)\beta^3/(\beta^5+1)\approx0.321\log_2M \\ \hline
S_n & \{-1,0,1\} & \beta^3=\beta+1 & 
(\log_\beta M)/(7+2\beta^2)\approx0.235\log_2M 
\end{array}
$$

If we want to compute a scalar multiple of a group element, e.g. a point $P$ 
on an elliptic curve, we can choose a representation $N=\sum_{j=0}^n x_jU_j$ of 
the scalar, compute $U_jP$, $0\le j\le n$, by using the recurrence of $U$ and 
finally $NP=\sum_{j=0}^n x_j(U_jP)$.
In the cases which we have considered, this amounts to $n+\|N\|_U$ 
additions (or subtractions).
Since $n\approx\log_\beta N$ is larger than $\|N\|_U$, the smallest number of 
additions is usually given by a $2$-expansion of minimal weight.
(We have $\log_{(1+\sqrt5)/2}N\approx 1.44\log_2 N$,
$\log_\beta N\approx 1.137\log_2 M$ for the Tribonacci number, 
$\log_\beta N\approx 2.465\log_2 N$ for the smallest Pisot number.)

If however we have to compute several multiples $NP$ with the same $P$ and
different $N\in\{-M,\ldots,M\}$, then it suffices to compute $U_jP$ for
$0\le j\le n\approx\log_\beta M$ once, and do $\|N\|_U$ additions for each $N$. 
Starting from $10$ multiples of the same $P$, the Fibonacci numeration system 
is preferable to base $2$ since $(1+10/5)\log_{(1+\sqrt5)/2}M\approx 
4.321\log_2 M<(1+10/3)\log_2 M$.
Starting from $20$ multiples of the same $P$, $S$-expansions of minimal weight
are preferable to the Fibonacci numeration system since 
$(1+20/(7+2\beta^2))\log_\beta M\approx 7.156\log_2 M<
7.202\log_2 M\approx(1+20/5)\log_{(1+\sqrt5)/2}M$.

\bibliography{minimal}
\end{document}